\documentclass[a4paper,fleqn]{cas-sc}

\usepackage[authoryear]{natbib}

\usepackage{subfigure}
\usepackage{graphicx}
\usepackage{amsmath}

\def\tsc#1{\csdef{#1}{\textsc{\lowercase{#1}}\xspace}}
\tsc{WGM}
\tsc{QE}

\begin{document}
\let\WriteBookmarks\relax
\def\floatpagepagefraction{1}
\def\textpagefraction{.001}

\shorttitle{}    

\shortauthors{Jiang}  

\title [mode = title]{Hierarchical Bidirectional Transition Dispersion Entropy-based Lempel-Ziv Complexity and Its Application in Fault-Bearing Diagnosis}  



%

\author[1]{Runze Jiang}[orcid=0000-0003-4106-4900]
\ead{23111517@bjtu.edu.cn}
\cormark[1]
\cortext[1]{Corresponding author}

\author[1]{Pengjian Shang}
\ead{pjshang@bjtu.edu.cn}

\affiliation[1]{organization={School of Mathematics and Statistics}, 
	addressline={Beijing Jiaotong University},
	city={Beijing 100044}, 
	country={China}}
















\begin{abstract}
Lempel-Ziv complexity (LZC) is a key measure for detecting the irregularity and complexity of nonlinear time series and has seen various improvements in recent decades. However, existing LZC-based metrics, such as Permutation Lempel-Ziv complexity (PLZC) and Dispersion-Entropy based Lempel-Ziv complexity (DELZC), focus mainly on patterns of independent embedding vectors, often overlooking the transition patterns within the time series. To address this gap, this paper introduces a novel LZC-based method called Bidirectional Transition Dispersion Entropy-based Lempel-Ziv complexity (BT-DELZC). Leveraging Markov chain theory, this method integrates a bidirectional transition network framework with DELZC to better capture dynamic signal information. Additionally, an improved hierarchical decomposition algorithm is used to extract features from various frequency components of the time series. The proposed BT-DELZC method is first evaluated through four simulated experiments, demonstrating its robustness and effectiveness in characterizing nonlinear time series. Additionally, two fault-bearing diagnosis experiments are conducted by combining the hierarchical BT-DELZC method with various classifiers from the machine learning domain. The results indicate that BT-DELZC achieves the highest accuracy across both datasets, significantly outperforming existing methods such as LZC, PLZC, and DELZC in extracting features related to fault bearings.

\end{abstract}




\maketitle

\section{Introduction}
\label{sec1}
Nonlinear time series analysis has been a important research area in the past few decades \citep{nd1,nd2,nd3}. It is dedicated to extracting valid information from complex signals and widely applied in many real-world scenarios including industrial engineering \citep{nd4,nd5}, biology \citep{nd6} and financial market \citep{nd7,nd8}. Several traditional methods, such as wavelet transform \citep{wavelet} or empirical mode decomposition \citep{emd}, are often used to characterize time series and gain some effect. However, these traditional metrics have some drawbacks, such as mode aliasing, high algorithmic complexity, and lack of interpretability, which can make them less effective when applied to real-world signal analysis. Due to this reason, discovering novel and effective methods to characterize nonlinear time series accurately have always been a hot topic in this research field.

Entropy is widely used to characterize the complexity of real-world signals due to its effectiveness in detecting nonlinearity and irregularity within time series. Since Shannon introduced the concept of entropy in 1948 \citep{shannon}, various forms of entropy have been developed to better analyze nonlinear time series. One such metric is Approximate Entropy (AE) \citep{AE}, proposed by Pincus, which assesses the degree of irregularity by calculating the difference between m-dimensional and m+1-dimensional phase space reconstruction vectors to reflect series self-similarity. However, AE is sensitive to the length of the signal. To address this issue, Richman introduced Sample Entropy (SE) \citep{SE}, which improves computational efficiency for long signals and is suitable for time series with missing data. In 2002, Bandt and Pompe proposed Permutation Entropy (PE) \citep{PE}, which measures the complexity of nonlinear time series using ordinal patterns and provides a framework for estimating the probability distribution of time series. Nonetheless, PE only considers the order of elements in embedding vectors and overlooks the amplitude information in the time series. To address this limitation, Rostaghi introduced Dispersion Entropy (DE) \citep{DE} in 2016. Unlike PE, DE incorporates both the mean and amplitude of signals, and research has demonstrated that DE provides more effective and accurate results in characterizing the properties of nonlinear time series across various application scenarios \citep{DE1,DE2,DE3}.

Lempel-Ziv complexity (LZC) \citep{LZC} is another widely used method for measuring the complexity of nonlinear time series and garnered attention in several studies \citep{LZC1,LZC2}. Unlike entropy methods, LZC focuses on identifying new patterns that emerge in a time series as the sequence progresses from the starting to the ending point. Theoretically, a more complex sequence will exhibit a higher number of unique patterns. The original LZC method encoded time series into binary form, resulting in patterns consisting only of 0 and 1, which can lead to a loss of critical information contained in the amplitude. To address this limitation, researchers have worked on enhancing the original LZC method with various coding approaches. In 2015, Bai introduced Permutation Lempel-Ziv complexity (PLZC) \citep{PLZC} by combining ordinal pattern methods with the original LZC approach. Unlike the binary patterns used in LZC, PLZC counts new ordinal patterns that gradually appear in the sequence, thus capturing more information about the order of time series points. However, similar to Permutation Entropy (PE), PLZC may still overlook amplitude information. To address this, other derivative methods such as Symbolic Lempel-Ziv complexity (SLZC) \citep{SLZC} and Dispersion Entropy-based Lempel-Ziv complexity (DELZC) \citep{DELZC} have been proposed and have shown promising results in analyzing different types of signals. Despite these advancements, the existing LZC methods primarily focus on static patterns in time series, failing to capture the dynamic and transitional information between embedding vectors.

Ordinal Pattern Transition Network (OPTN), introduced by Small \citep{OPTN}, is a state-of-the-art method for analyzing time series by transforming them into directed graphs based on transition patterns between adjacent embedding vectors. Unlike traditional methods, OPTN offers a novel perspective by creating a detailed representation of time series dynamics. It has several advantages: it identifies more patterns by counting the links between adjacent vectors, allowing for a finer division of embedding vectors, and it captures the dynamic structure between vectors, revealing potential transformation information in nonlinear time series. Recently, OPTN has been integrated with various entropy measures and complexity metrics to enhance the characterization of nonlinear time series. Zhang developed Transition Permutation Entropy and Transition Dissimilarity Measures within the OPTN framework \citep{TPE}. Huang combined OPTN with statistical complexity measures to capture the dynamical transitions in nonlinear time series \citep{Huang}. Ding introduced Link Dispersion Entropy by substituting the ordinal patterns in OPTN with dispersion patterns for feature extraction in rotating machinery \citep{LDE}. However, integrating the time series transition network approach with LZC methods remains an unexplored area.

In this paper, we present the first attempt to integrate transition network methods with LZC approaches. Based on DELZC, a transition network framework utilizing dispersion patterns is incorporated into the original algorithm. Instead of the unidirectional transitions (links) commonly used in most transition network methods, a weighted bidirectional transition framework is established to reveal the dynamic structure of time series more comprehensively. As a result, a novel LZC-based method, named Bidirectional Transition Dispersion Entropy-based Lempel-Ziv Complexity (BT-DELZC), is proposed to better extract features from nonlinear time series. Furthermore, the hierarchical decomposition model \citep{hier} proposed by Jiang et al. is appended to analyze time series across different frequency components. Experiments are first conducted on several simulated signals to discuss parameter settings and demonstrate the advantages of BT-DELZC. Subsequently, the method is applied to two real-world fault-bearing datasets to verify its effectiveness. The results show that BT-DELZC achieves the highest accuracy compared to previously proposed LZC, PLZC, and DELZC methods, thereby proving its efficiency and practicality for feature extraction from nonlinear time series.

The structure of this paper is outlined as follows: Section~\ref{sec2} introduces the BT-DELZC algorithm framework, along with a brief review of previous LZC methods and transition network approaches. Section~\ref{sec3} presents simulated experiments to demonstrate the advantages of BT-DELZC. In Section~\ref{sec4}, the effectiveness of the BT-DELZC method is verified using real-world fault-bearing datasets. Section~\ref{dis} performs the discussion about the motivations, principles and outlooks of the proposed BT-DELZC. Finally, Section~\ref{con} provides the conclusions of the paper.

\section{Methods}
\label{sec2}
\subsection{Lempel-Ziv complexity}
\label{sec2a}
LZC is a widely used method for measuring the complexity of nonlinear time series by focusing on the rate at which new patterns emerge within the sequences.  For a given univariate time series $X=\{x_{j}\}_{j=1,...,N}$, the steps to calculate LZC are as follows:

Step 1: Calculate the mean value of the time series $X$ as threshold and convert it to binary sequence $Z$ by Eq.~\eqref{eq1}:
  \begin{equation}
  	\label{eq1}
  	z_j= \begin{cases}1 & \text { if } x_j \geq \operatorname{mean}\left(X\right) \\ 0 & \text { if } x_j<\operatorname{mean}\left(X\right)\end{cases}
  \end{equation}
where $j=1,2,...,N$.

Step 2: Initialize the set $S=\{z_1\}$ and $Q=\{\}$. Set the complexity value $C(N)=1$ and the counting parameter $k=2$. 

Step 3: Set $Q=\{Q,z_k\}$. Then, combine $S$ with $Q$ and named the new set $SQ$. Next, remove the element in the last position of $SQ$ and gain the set $SQv$. Transform the parameter $k$ into $k+1$.

Step 4: Judge whether $Q$ is the subset of $SQv$. If so, return to Step 3 and repeat the instruction. If not, update $C(N)=C(N)+1$ and take Step 5. 

Step 5: Change $S$ into $SQ$ and reset $Q=\{\}$. Determine whether the parameter $k>N$. If so, take the Step 6 to calculate final LZC measure of the sequence. If not, return to Step 3.

Step 6: Calculate the final LZC measure by the following Eq.~\eqref{eq2} and Eq.~\eqref{eq3}:
 
\begin{equation}
	\label{eq2}
	C^*= \lim_{N \to \infty} C(N) \approx \frac{N}{\log_2 N}
\end{equation}

\begin{equation}
	\label{eq3}
	LZC = \frac{C(N)}{C^*}
\end{equation}

where $C^*$ is the normalized constant in order to alleviate the influence of sequence length $N$.

\subsection{Permutation Lempel-Ziv complexity and Dispersion entropy-based Lempel-Ziv complexity}
\label{sec2b}
As noted, while LZC is a useful indicator for measuring the complexity of time series, it only divides the sequence into two symbols, 0 and 1. Consequently, important information contained in the amplitude may be lost. Recently, several symbolization methods, such as PLZC and DELZC, have been proposed to enhance the effectiveness of the original LZC approach.

\subsubsection{Permutation Lempel-Ziv complexity}
\label{sec2b1}
The PLZC method transforms the binary 0-1 symbols into the ordinal patterns introduced by Bandt and Pompe[]. Unlike LZC, which operates on binary symbols, PLZC uses phase space reconstruction to capture more detailed information by embedding vectors. Given a univariate time series $X=\{x_{j}\}_{j=1,...,N}$, the first step involves reconstructing the series into a series of vectors  $x_{j}^{m,\tau}=\{x_{j},x_{j+\tau},...,x_{j+(m-1)\tau}\}\,,\, j=1,2,...,N-(m-1)\tau$, with an embedding dimension $m$ and time delay $\tau$. Each embedding vector is then transformed into an ordinal pattern based on the rank order of its $m$ elements.  For example, the rank order of the embedding vector $(x_{t}, x_{t+1}, x_{t+2},x_{t+3})=(9,4,7,3)$ is $(x_{t+3}<x_{t+1}<x_{t+2}<x_{t})$. resulting in the ordinal pattern $(4,2,3,1)$. For a chosen embedding dimension $m$, there are $m!$ possible ordinal patterns, represented by $\{\pi_{i}\}_{i=1,2,...,m!}$. The original time series $X$ is thus converted into a sequence of ordinal patterns $Z=\{\pi_{1}\pi_{2}\pi_{3}...\}$.  By applying the LZC algorithm to this ordinal pattern sequence, the PLZC value of the time series $X$ can be computed as follows:

\begin{equation}
	C^*= \lim_{N \to \infty} C(N) \approx \frac{N-(m-1)\tau}{\log_{m!} N-(m-1)\tau}
\end{equation}

\begin{equation}
	PLZC = \frac{C(N)}{C^*}
\end{equation}

Instead of using binary symbols like LZC, PLZC divides the time series into $m!$ distinct patterns through phase space reconstruction. This approach allows PLZC to capture more detailed information embedded in the phase space trajectories generated by the original sequence.

\subsubsection{Dispersion entropy-based Lempel-Ziv complexity}
\label{sec2b2}
DELZC, introduced by Li et al. in 2022, combines the DE proposed by Rostaghi with the LZC algorithm. In comparison with LZC and PLZC, DELZC incorporates both the means and amplitudes of the time series, thereby enhancing its ability to detect the structure of nonlinear time series. The detailed algorithm for DELZC applied to a time series $X=\{x_{j}\}_{j=1,...,N}$ is as follows:

Step 1: Map the elements of time series $X$ into $c$ classes using the normal cumulative distribution function (NCDF), resulting in a new sequence $Y={\{y_i\}}_{i=1,2,...,N}$. The NCDF is defined by the equation

\begin{equation}
	y_i=\frac{1}{\sigma \sqrt{2 \pi}} \int_{-\infty}^{x_i} e^{\frac{-(t-\mu)^2}{2 \sigma^2}} \mathrm{~d} t \text {, }
\end{equation} 

where $\mu$ and $\sigma$ represent the mean and standard deviation of $X$, respectively. The number of classes $c$ is the parameter chosen based on the specific characteristics of the time series. Subsequently, transform the sequence $Y$ into a symbolic sequence $Z$ using the equation $z_i=round(c \cdot y_i+0.5)$. Note that the elements of $Z$ are integers ranging from $1$ to $c$.

Step 2: Select the embedding dimension $m$ and time delay $\tau$. Reconstruct the sequence $Z$ into a series of vectors $z_i^{m,\tau}=\{z_i,z_i+\tau,...,z_{i+(m-1)\tau}\}$ for $i=1,2,...,N-(m-1)\tau$. Each vector is classified as a dispersion pattern, and a new symbolic sequence corresponding to these patterns is generated. This algorithm identifies a total of $c^m$ distinct patterns   

Step 3: Apply the LZC algorithm, as detailed in Section~\ref{sec2a}, to the new sequence to compute the complexity value $C(N)$. The DELZC equation is then defined as:

\begin{equation}
	C^*= \lim_{N \to \infty} C(N) \approx \frac{N-(m-1)\tau}{\log_{c^m} N-(m-1)\tau}
\end{equation}

\begin{equation}
	DELZC = \frac{C(N)}{C^*}
\end{equation}      

By incorporating specific amplitudes, DELZC outperforms both LZC and PLZC in feature extraction from nonlinear time series \citep{DELZC}.

\subsection{Bidirectional transition dispersion entropy-based Lempel-Ziv complexity}
\label{sec2c}

Existing LZC methods primarily focus on fixed patterns within sequences, which limits their ability to capture the dynamic transitions between vectors in the phase space trajectory. To address this, we propose the BT-DELZC method, which integrates transition network concepts with DELZC. The core idea of the time series transition network is to analyze the transformation patterns of adjacent embedding vectors, reflecting the principles of Markov chains in stochastic processes. This paper introduces a weighted bidirectional transition framework based on DELZC, accounting for both in-link and out-link transitions between neighboring vectors. The detailed algorithm is as follows:

Step 1: Given a univariate $X=\{x_{j}\}_{j=1,...,N}$, first obtain the symbolic sequence $Z$ by constructing dispersion patterns of the embedding vectors, as described in Section~\ref{sec2b2}. 

\begin{figure*}
	\centering
	\includegraphics[width=6in]{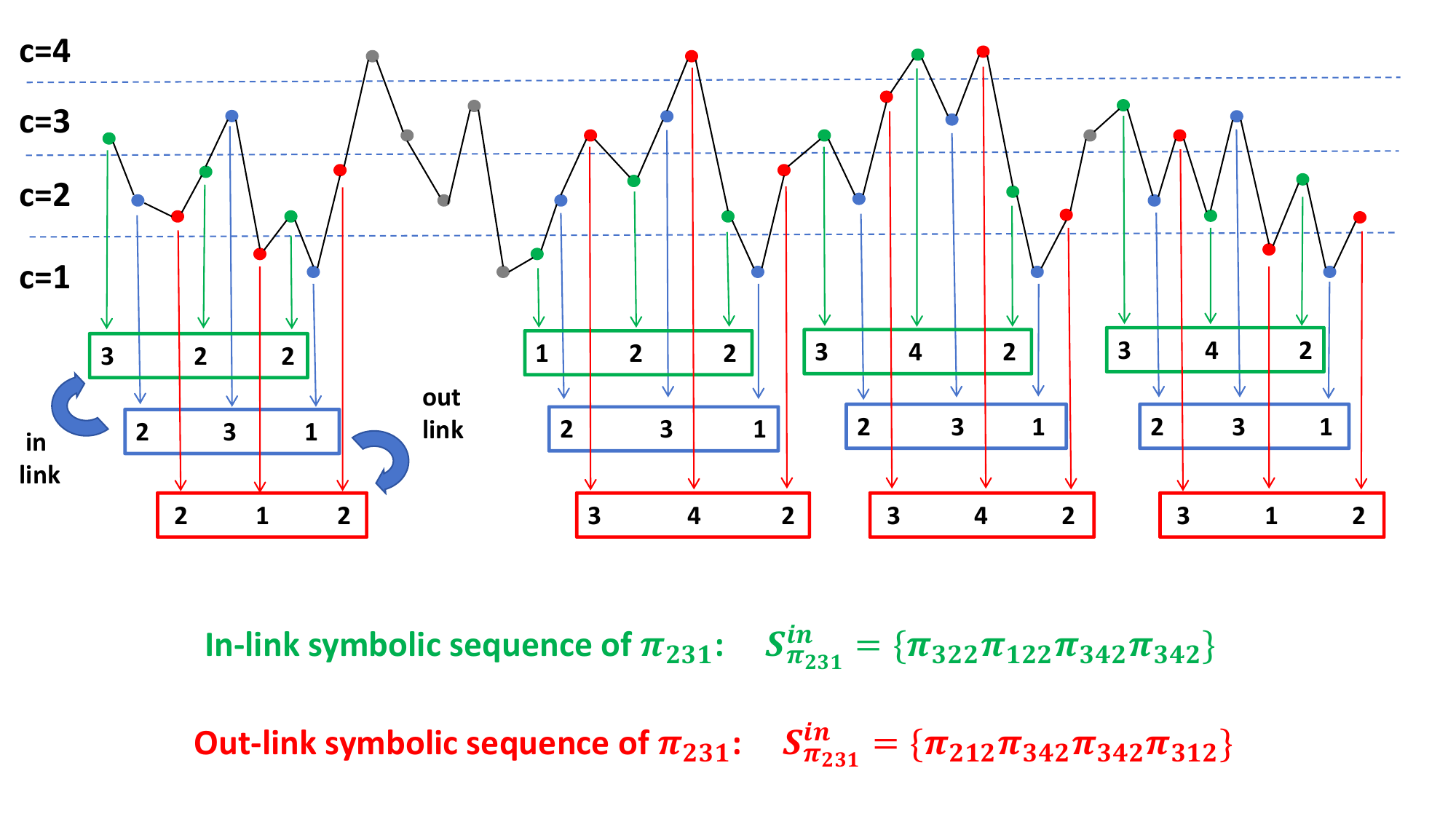}
	\caption{\label{fig1} An example of finding the In-link and Out-link sequences corresponding to the dispersion pattern $\pi_{231}$ }
\end{figure*}

Step 2: For each dispersion pattern ${\{\pi_{i}\}}_{i=1,2,...,c^m}$, identify its occurrences in the time series and extract the corresponding in-link and out-link sequences, denoted as  $S_{\pi_i}^{in}$ and $S_{\pi_i}^{out}$. To focus on transitions between different patterns, self-loops are excluded from this analysis. As a result, the possible symbols in $S_{\pi_i}^{in}$ and $S_{\pi_i}^{out}$ number $c^m-1$. (An example of this process is illustrated in Fig.~\ref{fig1}). To ensure the presence of both in-link and out-link vectors, dispersion patterns are selected within the range $i=2,3,...,N-1$, thereby excluding both the startpoints and endpoints of the time series.

Step 3: Compute the complexity values $C(N_{in})_{\pi_{i}}$ and $C(N_{out})_{\pi_{i}}$ for the sequences $S_{\pi_i}^{in}$ and $S_{\pi_i}^{out}$ for each dispersion pattern ${\{\pi_{i}\}}$ as outlined in Section~\ref{sec2b2}. Next, obtain $DELZC_{in}(\pi_{i})$ and $DELZC_{out}(\pi_{i})$ using the following formulas:

\begin{equation}
	C_{in}^*(\pi_{i})= \lim_{N_{in} \to \infty} C(N_{in})_{\pi_{i}} \approx \frac{N_{in}}{\log_{c^m-1} N_{in}}
\end{equation}

\begin{equation}
	C_{out}^*(\pi_{i})= \lim_{N_{out} \to \infty} C(N_{out})_{\pi_{i}} \approx \frac{N_{out}}{\log_{c^m-1} N_{out}}
\end{equation}

\begin{equation}
	DELZC_{in}(\pi_{i}) = \frac{C(N_{in})_{\pi_{i}}}{C_{in}^*(\pi_{i})}
\end{equation}  

\begin{equation}
	DELZC_{out}(\pi_{i}) = \frac{C(N_{out})_{\pi_{i}}}{C_{out}^*(\pi_{i})}
\end{equation} 

where $N_{in}$ and $N_{out}$ represents the lengths of $S_{\pi_i}^{in}$ and $S_{\pi_i}^{out}$, respectively. Note that $N_{in}$ and $N_{out}$ may differ due to varying numbers of self-loops in the series. The DELZC of $\pi_{i}$ is then defined as the mean of $DELZC_{in}(\pi_{i})$ and $DELZC_{out}(\pi_{i})$:

\begin{equation}
	DELZC_{\pi_{i}}=\frac{DELZC_{in}(\pi_{i})+DELZC_{out}(\pi_{i})}{2}
\end{equation} 

According to the algorithm above, a series of $DELZC_{\pi_{i}},i=1,2,...,c^m$ values are obtained.     

Step 4: Calculate the frequency of each dispersion pattern $\pi_{i}$ using the following equation:

\begin{equation}
	p(\pi_{i})=\frac{\|\,j:2\leq j\leq N-(m-1)\tau-1,\, z_{j}^{m,\tau}=type\,\{\pi_{i}\}\,\|}{N-(m-1)\tau-2}.
\end{equation}

Here, the frequency $p(\pi_{i})$ serves as the weight for each pattern. Finally, compute the sum of the products of these weights with their corresponding DELZC values. The BT-DELZC for the time series $X$ can then be expressed as:

\begin{equation}
	BT-DELZC(X)=\sum_{i=1}^{c^m} {p(\pi_{i}) DELZC(\pi_{i})}.
\end{equation} 

\begin{figure*}
	\centering
	\includegraphics[width=6in]{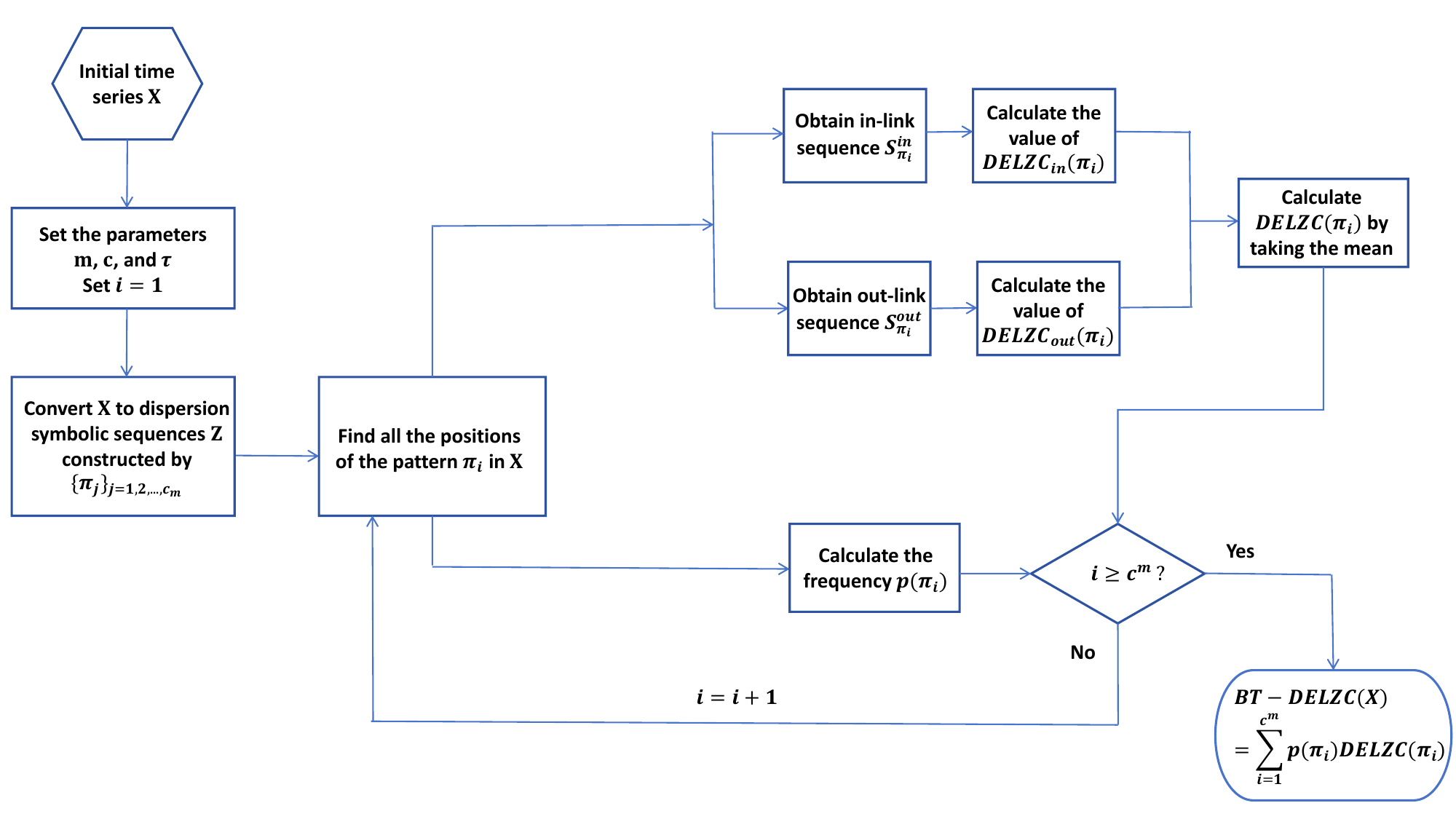}
	\caption{\label{fig2} The flow diagram of the procedure for calculating BT-DELZC of a given time series $X$ }
\end{figure*}

For a clearer understanding, the flow diagram for calculating BT-DELZC is illustrated in Fig.~\ref{fig2}. This diagram provides a comprehensive overview of the method, highlighting each step involved in the calculation process described above. 

\subsection{Hierarchical Bidirectional transition dispersion entropy-based Lempel-Ziv complexity}
\label{sec2d}

To more effectively characterize the features of nonlinear time series, this paper applies the hierarchical decomposition method. Compared to the multiscale method \citep{multiscale}, the hierarchical decomposition model can extract information from both low-frequency and high-frequency components of the time series. In the subsequent sections, an enhanced version of hierarchical decomposition, utilizing overlapped sliding windows as introduced by Zhao \citep{Zhao}, is employed. This approach helps mitigate the significant fluctuations in LZC values observed in high-frequency components. For a given time series $X=\{x_{i}\}_{i=1,...,N}$, the hierarchical decomposition involves two operators:

\begin{equation}
	\begin{aligned}
		& Q_0(x)=\frac{x(i)+x(i+1)}{2} \quad i=1,2,...,N-1 \\
		& Q_1(x)=\frac{x(i)-x(i+1)}{2} \quad i=1,2,...,N-1
	\end{aligned}
\end{equation}

In this equation, $Q_0(x)$ denotes the low-frequency operator, while $Q_1(x)$ denotes the high-frequency operator. During the calculations, the operators $Q_\varepsilon^K$ can be represented as a matrix given by:

\begin{equation}
	 \left[\begin{array}{cccccccc}
		\frac{1}{2} & \underbrace{0 \cdots 0}_{2^{K-1}-1} & \frac{(-1)^\varepsilon}{2} & 0 & \cdots & 0 & 0 & 0 \\
		0 & \frac{1}{2} & \underbrace{0 \cdots 0}_{2^{K-1}-1} & \frac{(-1)^\varepsilon}{2} & \cdots & 0 & 0 & 0 \\
		\cdots & \cdots & \cdots & \cdots & \cdots & \cdots & \cdots & \cdots \\
		0 & 0 & 0 & 0 & \cdots & \frac{1}{2} & \underbrace{0 \cdots 0}_{2^{K-1}-1} & \frac{(-1)^\varepsilon}{2}
	\end{array}\right]
\end{equation}

where $K$ represents the hierarchical layer and $\varepsilon \in \{0,1\}$ denotes the frequency indicator. As the number of decomposition layers $K$ increases, the scale of $Q_\varepsilon^K$ becomes $N-2^K+1 \times N-2^{K-1}+1$. Depending on the requirements, a K-dimensional indicator vector $(\lambda_1,\lambda_2,...,\lambda_K) \in \{0,1\}$ can be selected. The number of frequency components $e$ can then be represented as

\begin{equation}
	e=\sum_{j=1}^{K} 2^{K-j} \lambda_j
\end{equation}

Finally, the $e_{th}$ hierarchical components of time series $X$ can be computed by

\begin{equation}
	X_{K,e}=Q_{\lambda_K}^K Q_{\lambda_{K-1}}^{K-1} \ldots Q_{\lambda_2}^2 Q_{\lambda_1}^1 \cdot X 
\end{equation}

\begin{figure}
	\centering
	\includegraphics[width=5in]{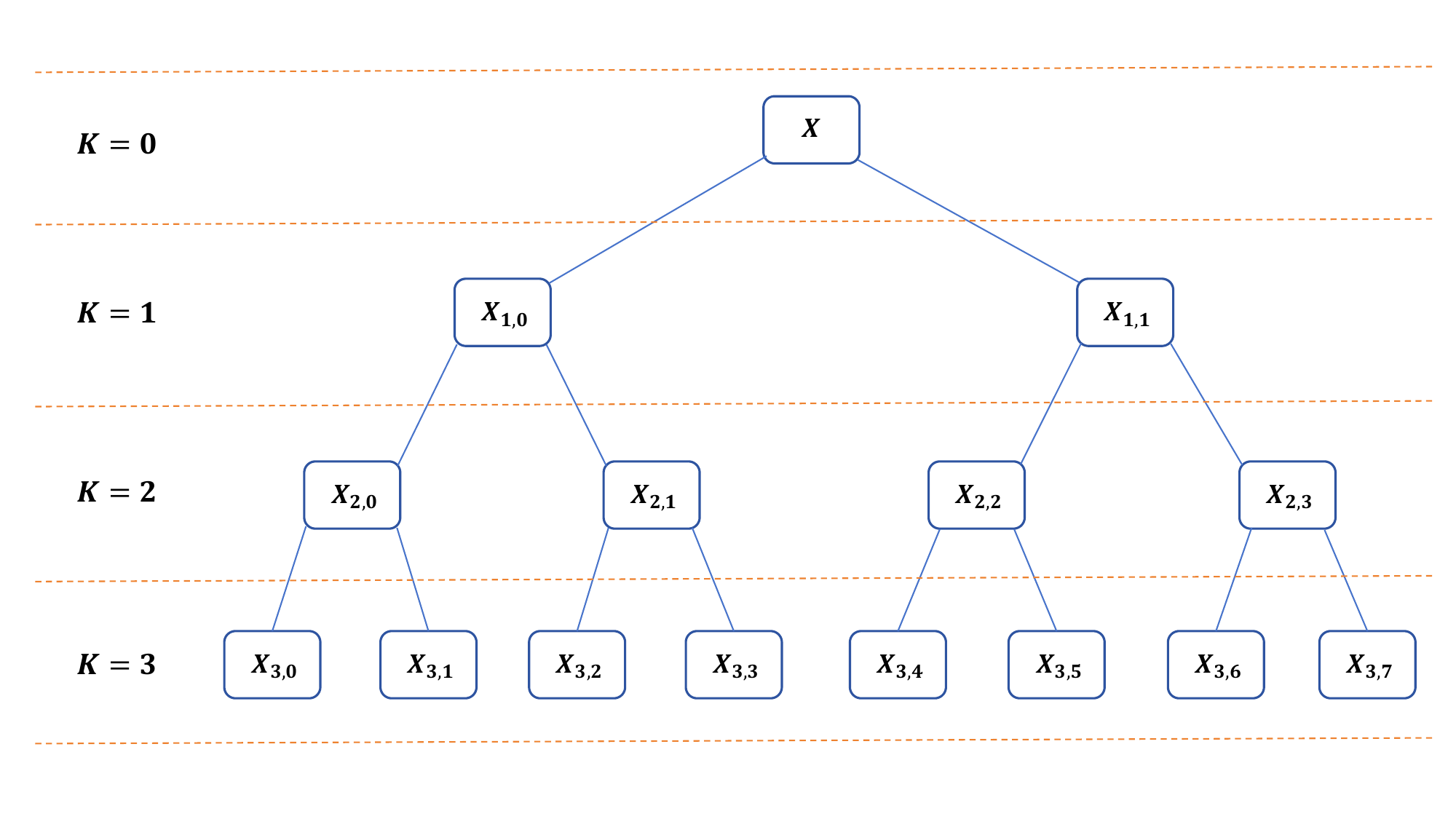}
	\caption{\label{fig3} The hierarchy diagram of $K=3$. }
\end{figure}

Different indicator vectors correspond to distinct frequency components. For the $K_{th}$ layer, there are a total of $2^K$ components. Consequently, $\sum_{j=1}^{K} 2^j$ components including the original time series $X$, can be obtained. Fig.~\ref{fig3} illustrates the hierarchy diagram of $K=3$ to aid in understanding. By calculating the BT-DELZC values for these different frequency components, multiple features of a given nonlinear time series can be extracted. 

\section{Experiments in Simulated Data}
\label{sec3}

\subsection{Parameter analysis}  
\label{sec3a}   

Before conducting the experiments, this subsection will discuss the selection of parameters such as time series length $N$, embedding dimension $m$, classification number $c$, and time delay $\tau$ to insure the accuracy and interpretability of the models. The embedding dimension $m$ is typically set to $m\geq3$. If $m$ is too small, the phase space trajectory may not be fully opened, making it difficult to extract sufficient information from the time series. Conversely, if $m$ is too large, the results will be susceptible to noise and the computational costs may increase. Due to the abundant dispersion patterns in the BT-DELZC algorithms, an embedding dimension of 3 or 4 is recommended. 
   
\begin{figure}
	\centering
	\includegraphics[width=5in]{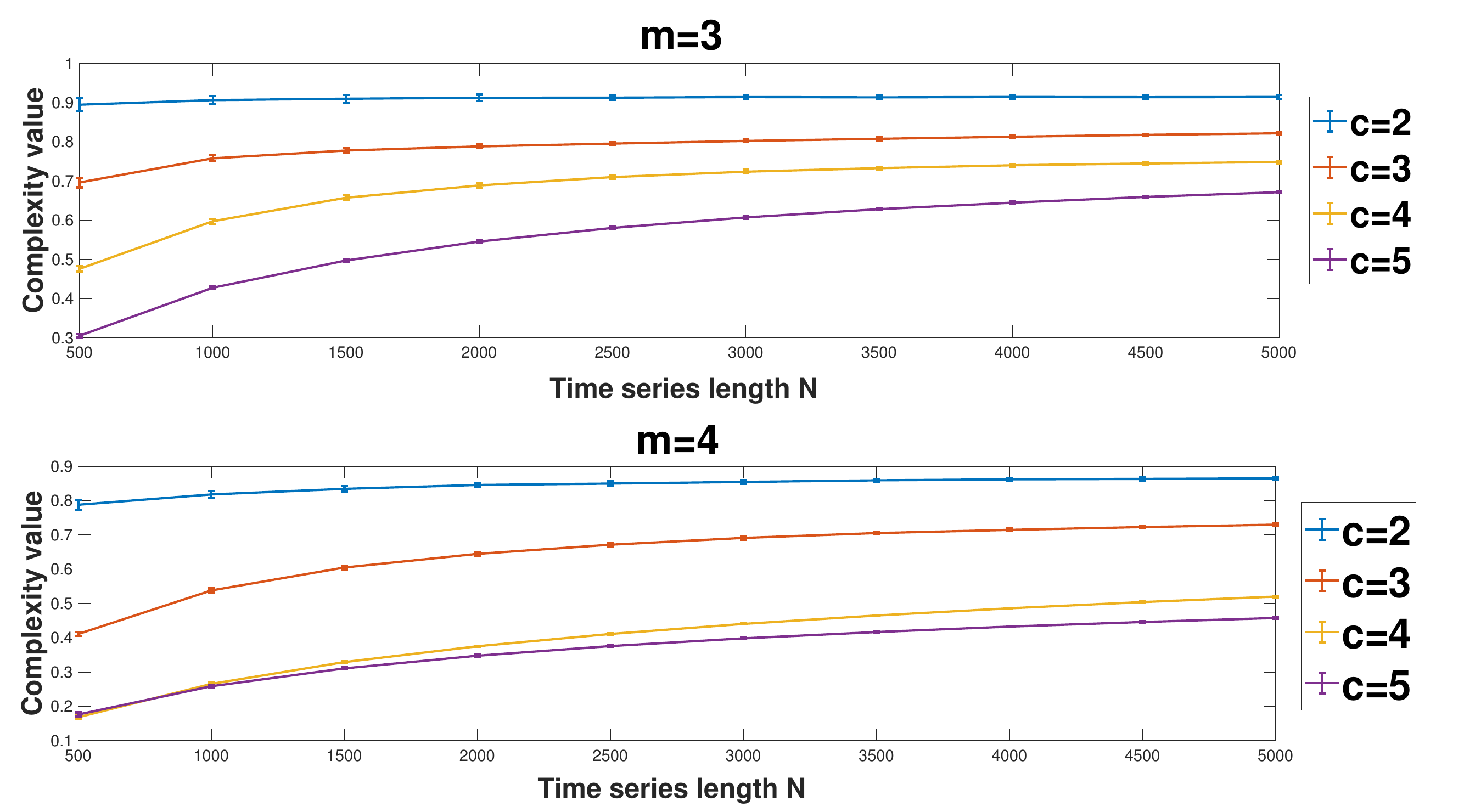}
	\caption{\label{fig4} The BT-DELZC results for $m=3$ and $m=4$ with varying time series length $N$ and different classification number $c$ }
\end{figure}

Subsequently, Gaussian white noise is used for further discussion. Fig.~\ref{fig4} shows the BT-DELZC results for $m=3$ and $m=4$ with varying time series length $N$ and different classification number $c$. When $c=2$, the results is robustness to $N$, but may lose information due to insufficient patterns. For $c$ equals to 3, 4, or 5, more information can be extracted from the series. However, the sequence should not be too short, as indicated in the figure, since the BT-DELZC values increase with $N$. It is advisiable to choose the time series $N$ such that the results approach their maximum for different $m$ and $c$. In conclusion, $N$ is recommended to be greater than 2000 when $c\geq3$ to ensure more accurate results.

Finally, for the time delay $\tau$, it is recommended to choose $\tau \geq 2$ to prevent overlap of adjacent vectors in the transition networks.

\subsection{Robustness Analysis}
\label{sec3b}

Robustness is a crucial factor in evaluating the quality of an algorithm. In this section, we conduct two simulated experiments to assess the robustness of the proposed BT-DELZC method in comparison with previous methods.

\begin{figure}
	\centering
	\includegraphics[width=5in]{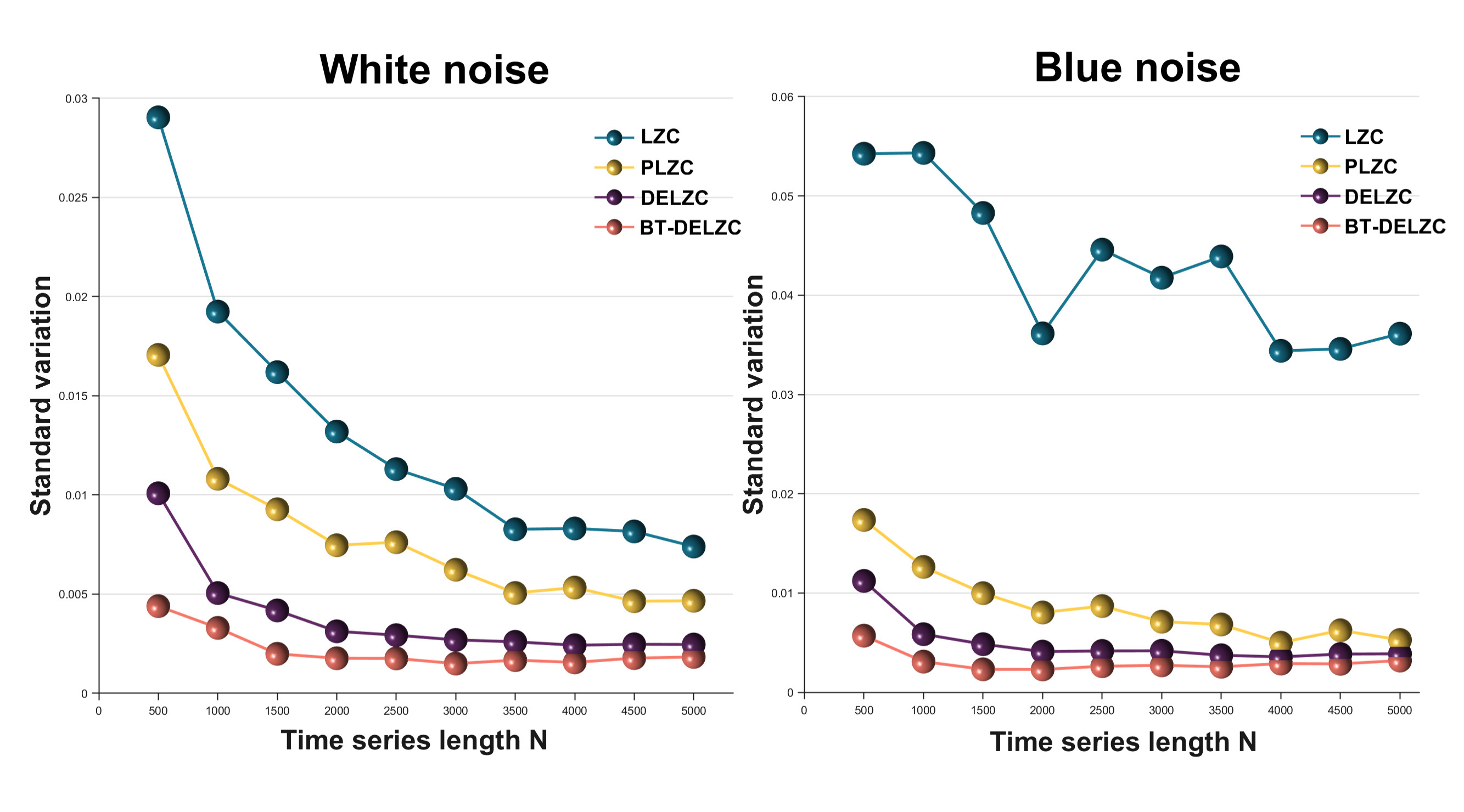}
	\caption{\label{fig5} The standard deviations of BT-DELZC and three previous methods (LZC, PLZC, DELZC) with different time series length $N$)}
\end{figure}

In the first experiment, we examine random error by calculating the standard deviation across repeated samples. Gaussian noise, including white noise and blue noise, of varying lengths from 
$N=500$ to $N=5000$, is used. For each length and type of noise, 100 samples are generated, and the standard deviations of BT-DELZC and three previous methods (LZC, PLZC, DELZC) are presented in Fig.~\ref{fig5}. The parameters for LZC, PLZC, and DELZC are set to 
$m=4$, $c=4$ and $\tau=2$. The results show that BT-DELZC consistently has the lowest standard deviation across all time series lengths and noise types, demonstrating the robustness and low-error performance of the proposed method.

\begin{figure*}
	\centering
	\includegraphics[width=6.5in]{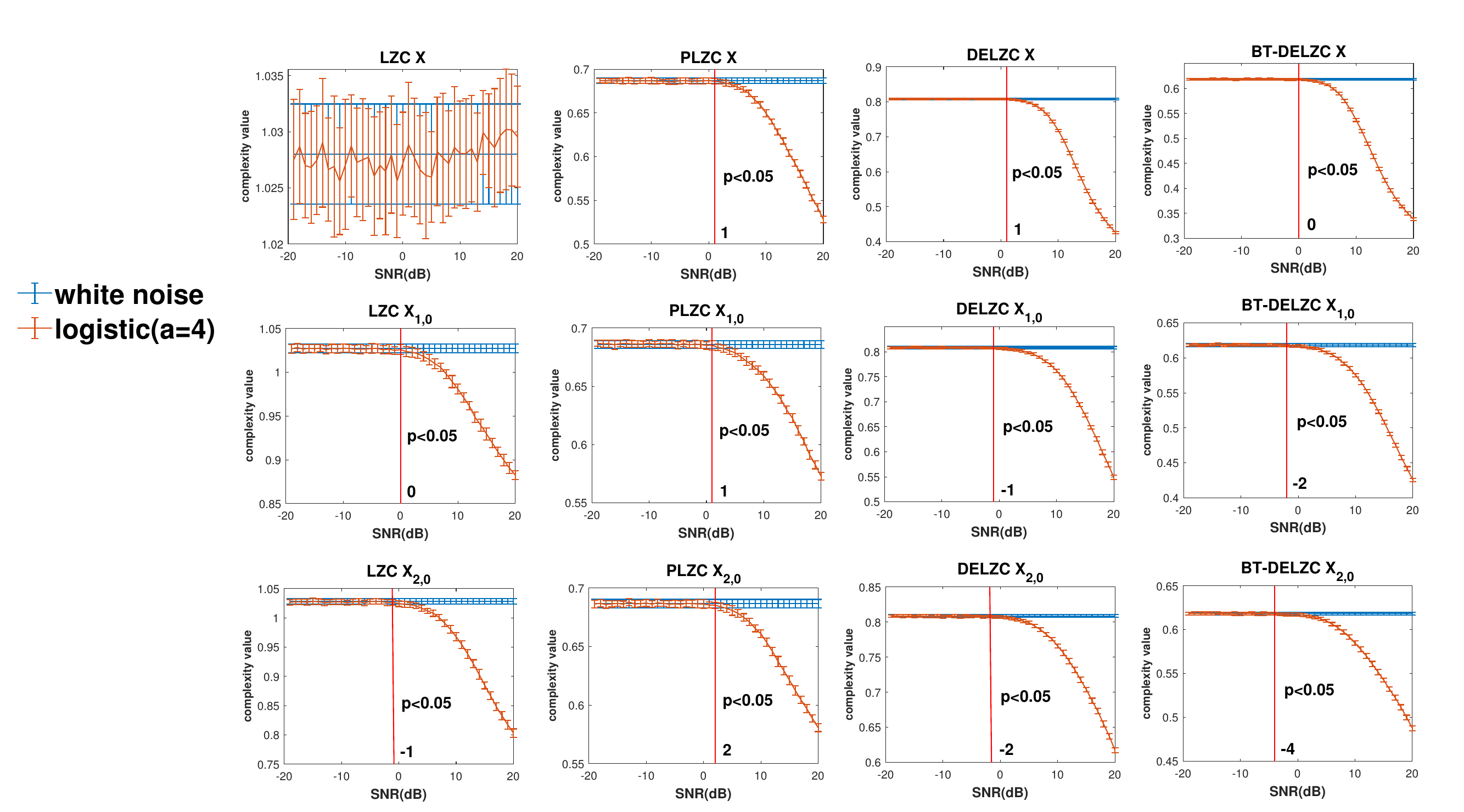}
	\caption{\label{fig6} The errorbar plots of the whitenoise and the fully chaotic logistic map (a=4) under the varying SNR, including four methods (LZC, PLZC, DELZC, BT-DELZC) and three hierarchies ($X$, $X_{1,0}$, $X_{2,0}$). The minimum SNR for $p<0.05$ is marked by red line in each cases.}
\end{figure*}

In the second experiment, the robustness of the algorithms to noise are analyzed and compared. Distinguishing chaos from stochastic noise is a common challenge in nonlinear time series research due to their similar characteristics in time domain waveforms. The logistic map is used to conduct this experiment, described by the formula $x_{i+1}=ax_{i}(1-x_{i})$, where $a$ is the chaotic parameter varying from $[3,4]$. For this experiment, the logistic sequence with $a=4$ is generated to achieve a fully chaotic state. Then, the independent white noise is added to the fully chaotic logistic map, varying the signal-to-noise ratio (SNR) from -19dB to 20dB in 1dB increments. The robustness of the four methods is compared using three hierarchies: the original time series $X$, $X_{1,0}$ and $X_{2,0}$. For each SNR, results from 50 repetitions with a time series length $N=10000$ are analyzed. The parameter of LZC, PLZC and DELZC are set to $m=4$, $c=4$ and $\tau=2$, with the results presented in Fig.~\ref{fig6}. Error bars are ploted for each case, and the Wilcoxon rank sum test is performed to determine whether the logistic map with added noise differs significantly from the generated white noise. The results indicate that LZC cannot distinguish between white noise and the fully chaotic logistic map when using the original sequence $X$.  For other scenarios, the red line indicates the minimum SNR where statistical significance $p<0.05$ is achieved. Statistical significance around the transition points is detailed in Table.~\ref{tab1}. The results demonstrate that BT-DELZC has the lowest statistical significance in most of the conditions compared to the other methods. Overall, BT-DELZC exhibits superior robustness and performance in distinguishing between fully chaotic logistic maps and white noise compared to LZC, PLZC, and DELZC.

\begin{table}
	\centering
	\caption{The statictical significance results of Wilcoxon rank sum test for four methods (LZC, PLZC, DELZC, BT-DELZC) under three hierarchies ($X$, $X_{1,0}$, $X_{2,0}$)}
	\resizebox{0.7\linewidth}{!}{\begin{tabular}{cccc|rrrrr}
			\toprule
			
			\multicolumn{2}{c}{\multirow{5}{*}{\textbf{Hierarchy $X$}}} & \multicolumn{2}{c}{\multirow{2}{*}{\textbf{LZC metric}}} &
			\multicolumn{5}{c}{\textbf{Signal-to-noise ratios (SNR)}} \\
			\cmidrule{5-9} \multicolumn{4}{c}{} &
			\multicolumn{1}{c}{5} & \multicolumn{1}{c}{4} &
			\multicolumn{1}{c}{3} & \multicolumn{1}{c}{2} &
			\multicolumn{1}{c}{1}  \\
			\cmidrule{3-9} \multicolumn{2}{c}{} &
			\multicolumn{2}{c}{LZC} & \multicolumn{1}{c}{**} & \multicolumn{1}{c}{**} & \multicolumn{1}{c}{**} & \multicolumn{1}{c}{**} & \multicolumn{1}{c}{**} \\
			\cmidrule{3-4} \multicolumn{2}{c}{} &
			\multicolumn{2}{c}{PLZC} & \multicolumn{1}{c}{2.96e-15} & \multicolumn{1}{c}{3.70e-10} & \multicolumn{1}{c}{6.79e-05} & \multicolumn{1}{c}{0.024} & \multicolumn{1}{c}{>0.05}  \\
			\cmidrule{3-4} \multicolumn{2}{c}{} &
			\multicolumn{2}{c}{DELZC} & \multicolumn{1}{c}{6.96e-18} & \multicolumn{1}{c}{1.72e-16} & \multicolumn{1}{c}{2.22e-08} & \multicolumn{1}{c}{4.47e-04} & \multicolumn{1}{c}{>0.05}  \\
			\cmidrule{3-4} \multicolumn{2}{c}{} &
			\multicolumn{2}{c}{\textbf{BT-DELZC}} & \multicolumn{1}{c}{\textbf{7.07e-18}} & \multicolumn{1}{c}{\textbf{7.07e-18}} & \multicolumn{1}{c}{\textbf{4.90e-13}} & \multicolumn{1}{c}{\textbf{2.29e-07}} & \multicolumn{1}{c}{\textbf{0.0071}}  \\
			
			\midrule
			
			\multicolumn{2}{c}{\multirow{5}{*}{\textbf{Hierarchy $X_{1,0}$}}} & \multicolumn{2}{c}{\multirow{2}{*}{\textbf{LZC metric}}} &
			\multicolumn{5}{c}{\textbf{Signal-to-noise ratios (SNR)}} \\
			\cmidrule{5-9} \multicolumn{4}{c}{} &
			\multicolumn{1}{c}{3} & \multicolumn{1}{c}{2} &
			\multicolumn{1}{c}{1} & \multicolumn{1}{c}{0} &
			\multicolumn{1}{c}{-1}  \\
			\cmidrule{3-9} \multicolumn{2}{c}{} &
			\multicolumn{2}{c}{LZC} & \multicolumn{1}{c}{2.08e-07} & \multicolumn{1}{c}{3.89e-05} & \multicolumn{1}{c}{0.0017} & \multicolumn{1}{c}{>0.05} & \multicolumn{1}{c}{**} \\
			\cmidrule{3-4} \multicolumn{2}{c}{} &
			\multicolumn{2}{c}{PLZC} & \multicolumn{1}{c}{2.33e-06} & \multicolumn{1}{c}{3.13e-04} & \multicolumn{1}{c}{>0.05} & \multicolumn{1}{c}{**} & \multicolumn{1}{c}{**}  \\
			\cmidrule{3-4} \multicolumn{2}{c}{} &
			\multicolumn{2}{c}{DELZC} & \multicolumn{1}{c}{1.91e-11} & \multicolumn{1}{c}{5.27e-09} & \multicolumn{1}{c}{5.64e-04} & \multicolumn{1}{c}{0.044} & \multicolumn{1}{c}{>0.05}  \\
			\cmidrule{3-4} \multicolumn{2}{c}{} &
			\multicolumn{2}{c}{\textbf{BT-DELZC}} & \multicolumn{1}{c}{\textbf{1.64e-04}} & \multicolumn{1}{c}{\textbf{3.96e-12}} & \multicolumn{1}{c}{\textbf{7.55e-06}} & \multicolumn{1}{c}{\textbf{0.0076}} & \multicolumn{1}{c}{\textbf{0.0047}}  \\
			
			\midrule
			
			\multicolumn{2}{c}{\multirow{5}{*}{\textbf{Hierarchy $X_{2,0}$}}} & \multicolumn{2}{c}{\multirow{2}{*}{\textbf{LZC metric}}} &
			\multicolumn{5}{c}{\textbf{Signal-to-noise ratios (SNR)}} \\
			\cmidrule{5-9} \multicolumn{4}{c}{} &
			\multicolumn{1}{c}{1} & \multicolumn{1}{c}{0} &
			\multicolumn{1}{c}{-1} & \multicolumn{1}{c}{-2} &
			\multicolumn{1}{c}{-3}  \\
			\cmidrule{3-9} \multicolumn{2}{c}{} &
			\multicolumn{2}{c}{LZC} & \multicolumn{1}{c}{1.52e-05} & \multicolumn{1}{c}{5.37e-04} & \multicolumn{1}{c}{>0.05} & \multicolumn{1}{c}{**} & \multicolumn{1}{c}{**} \\
			\cmidrule{3-4} \multicolumn{2}{c}{} &
			\multicolumn{2}{c}{PLZC} & \multicolumn{1}{c}{>0.05} & \multicolumn{1}{c}{**} & \multicolumn{1}{c}{**} & \multicolumn{1}{c}{**} & \multicolumn{1}{c}{**}  \\
			\cmidrule{3-4} \multicolumn{2}{c}{} &
			\multicolumn{2}{c}{DELZC} & \multicolumn{1}{c}{1.63e-05} & \multicolumn{1}{c}{0.0064} & \multicolumn{1}{c}{0.0045} & \multicolumn{1}{c}{>0.05} & \multicolumn{1}{c}{**}  \\
			\cmidrule{3-4} \multicolumn{2}{c}{} &
			\multicolumn{2}{c}{\textbf{BT-DELZC}} & \multicolumn{1}{c}{\textbf{1.30e-05}} & \multicolumn{1}{c}{\textbf{0.012}} & \multicolumn{1}{c}{\textbf{0.002}} & \multicolumn{1}{c}{\textbf{0.045}} & \multicolumn{1}{c}{\textbf{0.031}}  \\

			\bottomrule
			\label{tab1}
	\end{tabular}}
\end{table}

\subsection{Chirp Signal Experiment}
\label{sec3c}
\begin{figure}
	\centering
	\includegraphics[width=3.4in]{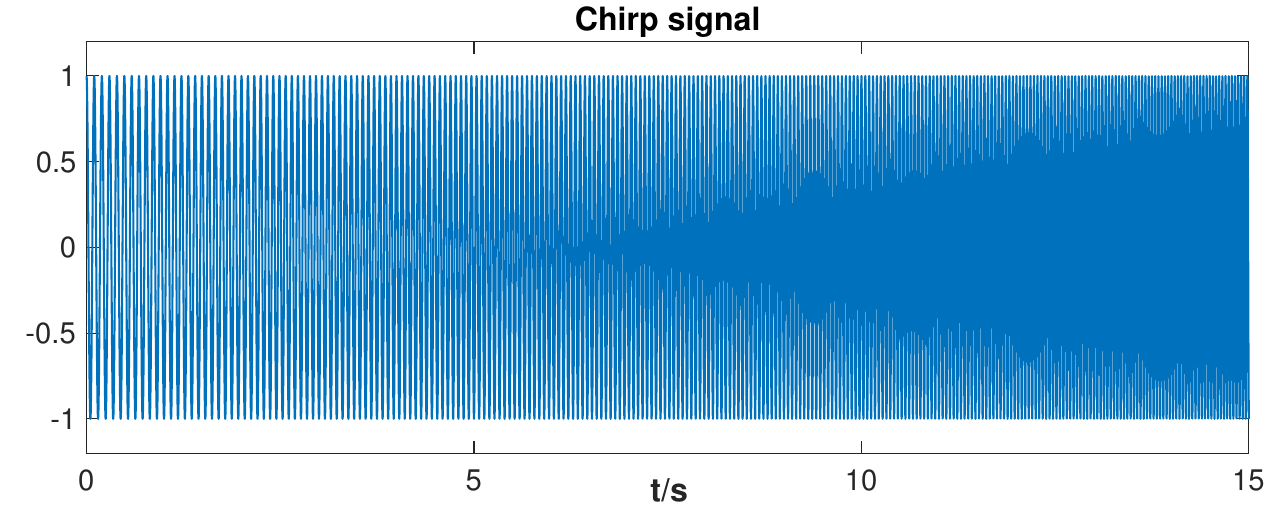}
	\caption{\label{fig7} The generated chirp signal with the frequency ranges from 10 Hz to 25 Hz. The sampling frequency is 1000 Hz. }
\end{figure}

The chirp signal \citep{chirp} is frequently used in simulations to assess the capability of algorithms in detecting signal characteristics due to its frequency variation over time. In this section, we generate a swept-frequency cosine chirp signal, where the instantaneous frequency is given by  $f(t)=f_0+u_0 t$, representing a linear increase. The equation for the chirp signal is then expressed as follows:
 
\begin{equation}
 	x(t)=exp\,(j2\pi(f_0t+\frac{1}{2}u_0 t))
\end{equation}

where $f_0$ is the initiation frequency and $u_0$ is the modulation frequency, defined as the frequency increase per second. In this experiment, the frequency ranges from 10 Hz to 25 Hz over a duration of 15 seconds, with a sampling frequency of 1000 Hz. This results in a total of 15,000 sample points, and the generated chirp signal is depicted in Fig.~\ref{fig7}.

\begin{figure}
	\centering
	\includegraphics[width=3.5in]{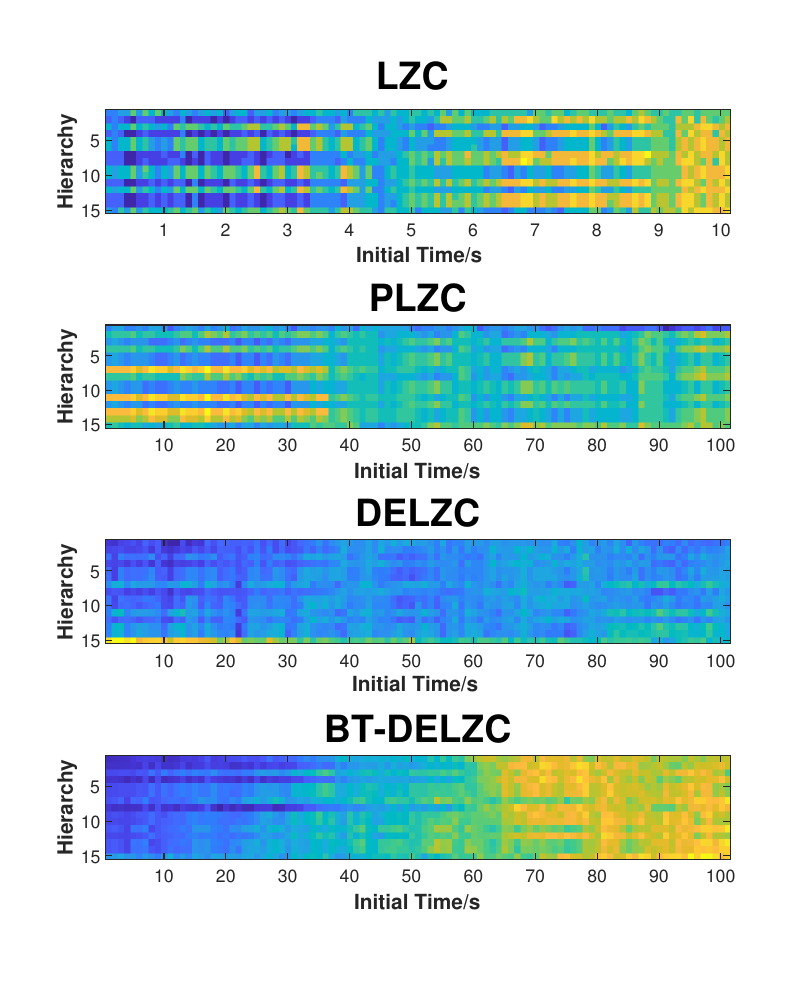}
	\caption{\label{fig8} The values of LZC, PLZC, DELZC, and BT-DELZC with the frequency changes in the chirp signals under 15 hierarchies.  }
\end{figure}

In our experiments, a sliding window of length 5000 is utilized. The window shifts 100 sample points to the right with each iteration, resulting in 101 sequences from the initial to the final point. For each sequence, hierarchical decomposition with $K=3$ is performed, leading to a total 15 hierarchies. The values of LZC, PLZC, DELZC, and BT-DELZC are computed for these hierarchies, using parameters $m=4$, $c=4$ and $\tau=2$. Fig.~\ref{fig8} presents a heatmap that illustrates the variation trends of these four methods. The results demonstrate that BT-DELZC effectively characterizes frequency changes, while the other three methods, particularly PLZC and DELZC, perform less effectively in detecting increasing changes. This comparison highlights the superior efficiency of the proposed BT-DELZC algorithm.

 \subsection{Mix Process Experiment}   
 \label{sec3d}
The mix process \citep{mix} generates a sequence that gradually transitions from a stochastic to a periodic time series. This process is useful for evaluating the ability of an algorithm to measure signal complexity. The formulation of the mix process is expressed as follows:

\begin{equation}
	MIX=(1-U)X+UY
\end{equation}  

\begin{figure}
	\centering
	\includegraphics[width=3.4in]{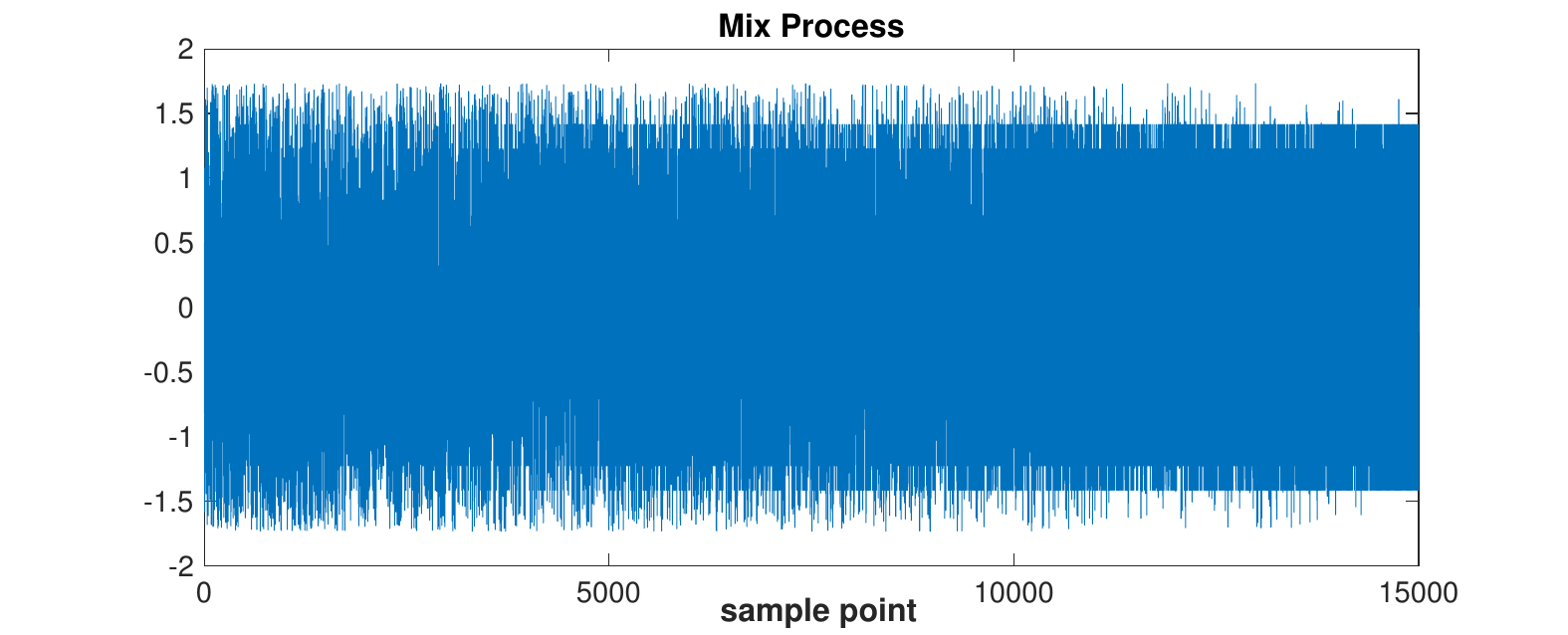}
	\caption{\label{fig9} The generated mix process with the sampling frequency selected as 1000 Hz.}
\end{figure}

In the formulation, $X$ represents a sinusoidal periodic series defined by $X_k=\sqrt{2} \sin \frac{2 \pi k}{12}$. $Y$ consists of uniformly distributed random variables over the interval $[-\sqrt{3}, \sqrt{3}]$. The parameter $U$ is a random variable with a two-point distribution: it equals 1 with probability $p$ and equals 0 with probability $1-p$. The value of $p$ linearly decreases from 0.99 at the initial sample point to 0.01 at the final sample point. In this experiment, the sampling frequency is set at 1000 Hz and the time duration is 15 seconds. The generated mix process is illustrated in Fig.~\ref{fig9}. The sliding window used has a length of 5 seconds and shifts 100 sample points to the right each time.

\begin{figure}
	\centering
	\includegraphics[width=4in]{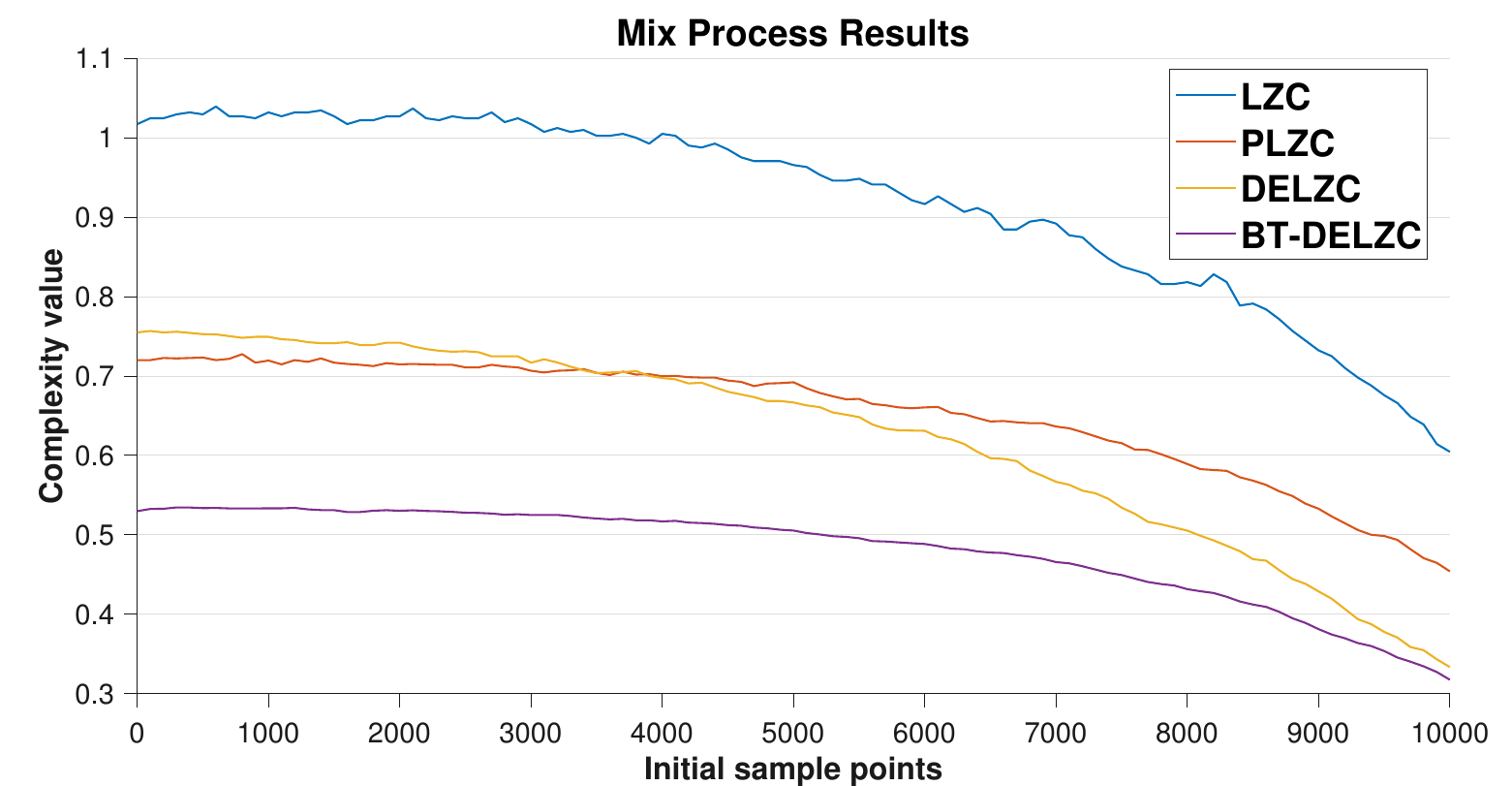}
	\caption{\label{fig10} The values of LZC, PLZC, DELZC, and BT-DELZC with the characteristic varying in the mix process. }
\end{figure}

The corresponding values of the four LZC-based methods (LZC, PLZC, DELZC, BT-DELZC) at various time points are shown in Fig.~\ref{fig10}. The parameters used are $m=4$, $c=4$ and $\tau=2$. The results indicate that all four LZC-based methods are capable of detecting the transition from stochastic to periodic sequences. Notably, the decline observed with the proposed BT-DELZC method is the smoothest among the four, highlighting its robustness and accuracy in capturing the characteristics of nonlinear time series.

\section{The Real-world Fault Bearing Diagnosis Experiments} 
\label{sec4}  
\subsection{Descriptions of Experimental Process}
\label{sec4a}
Fault bearing diagnosis is a critical issue in industrial engineering due to its essential role in maintaining the safe and efficient operation of rotating machinery. Consequently, numerous studies have been conducted to improve the accuracy of bearing diagnosis \citep{bear1,bear2,bear3,bear4}. In this paper, the BT-DELZC metric, combined with hierarchical decomposition, is applied to extract features from real-world fault bearing signals. To evaluate the effectiveness of the BT-DELZC measure, two publicly available datasets—fault bearing signals from Paderborn University (PU) \citep{PU} and Case Western Reserve University (CWRU) \citep{CWRU} —are utilized. The classification results of LZC, PLZC, DELZC, and BT-DELZC are collected and compared. The procedures for the fault bearing diagnosis experiments are illustrated in Fig.~\ref{fig11} and described as follows:

\begin{figure}
	\centering
	\includegraphics[width=4.5in]{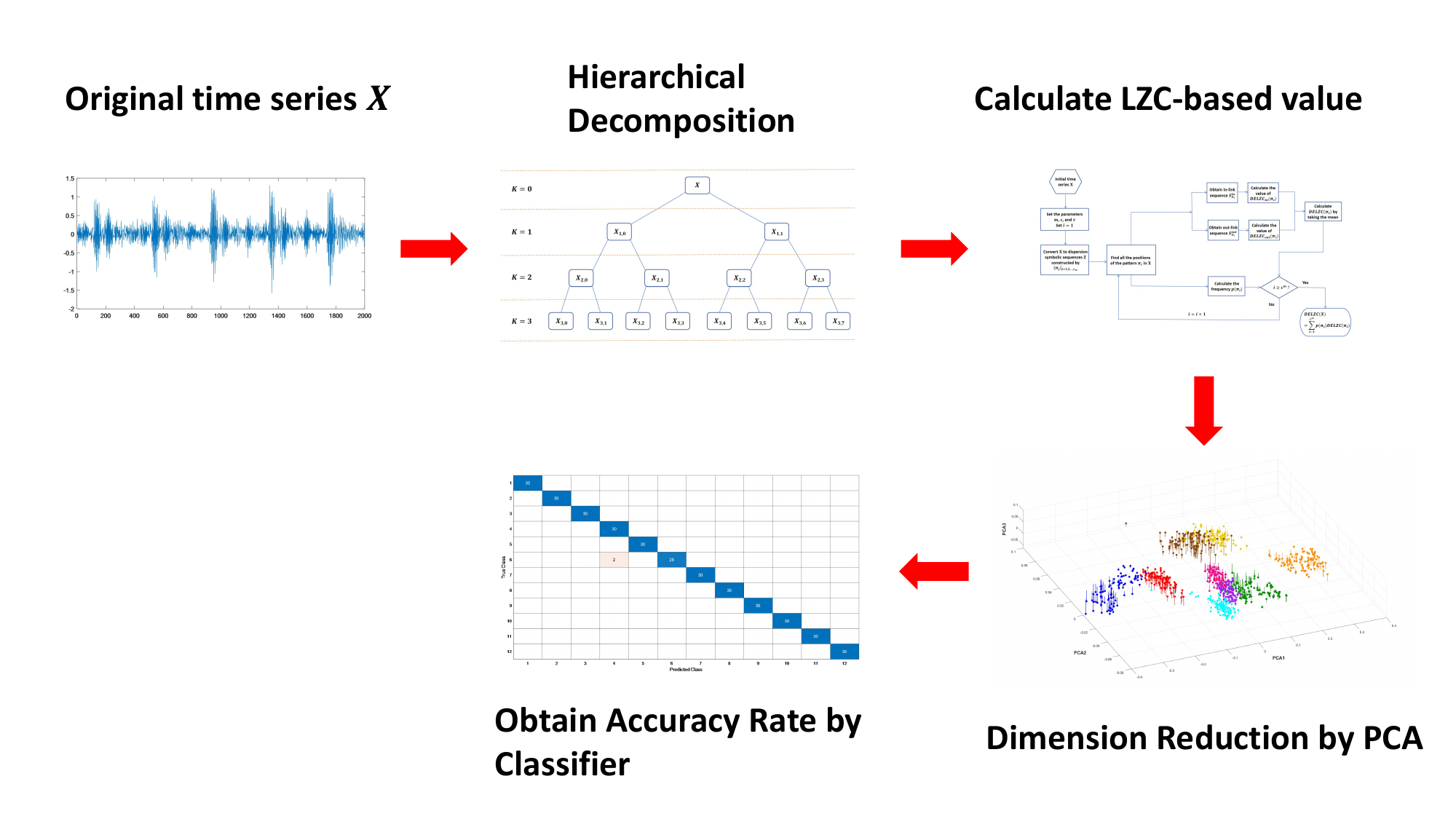}
	\caption{\label{fig11} The procedures for the fault bearing diagnosis in the following experiments. }
\end{figure}

Step 1: Obtain the original fault bearing signals from various fault types.

Step 2: Perform hierarchical decomposition on each sample according to the selected layer $K$.

Step 3: Compute the LZC-based measures across all hierarchies to extract multiple features for each sample.

Step 4: Apply Principal Component Analysis (PCA) to reduce the dimensionality of the features, emphasizing important characteristics and minimizing redundancy.

Step 5: After PCA, obtain the features for each sample and split them into training and test sets. Input the training set features into machine learning classifiers to train the model. Finally, use the well-trained classifier to classify the fault bearing types in the test set and calculate the accuracy rate.

\subsection{Paderborn University Fault Bearing Diagnosis}  
\label{sec4b}
\begin{figure}
	\centering
	\includegraphics[width=3.4in]{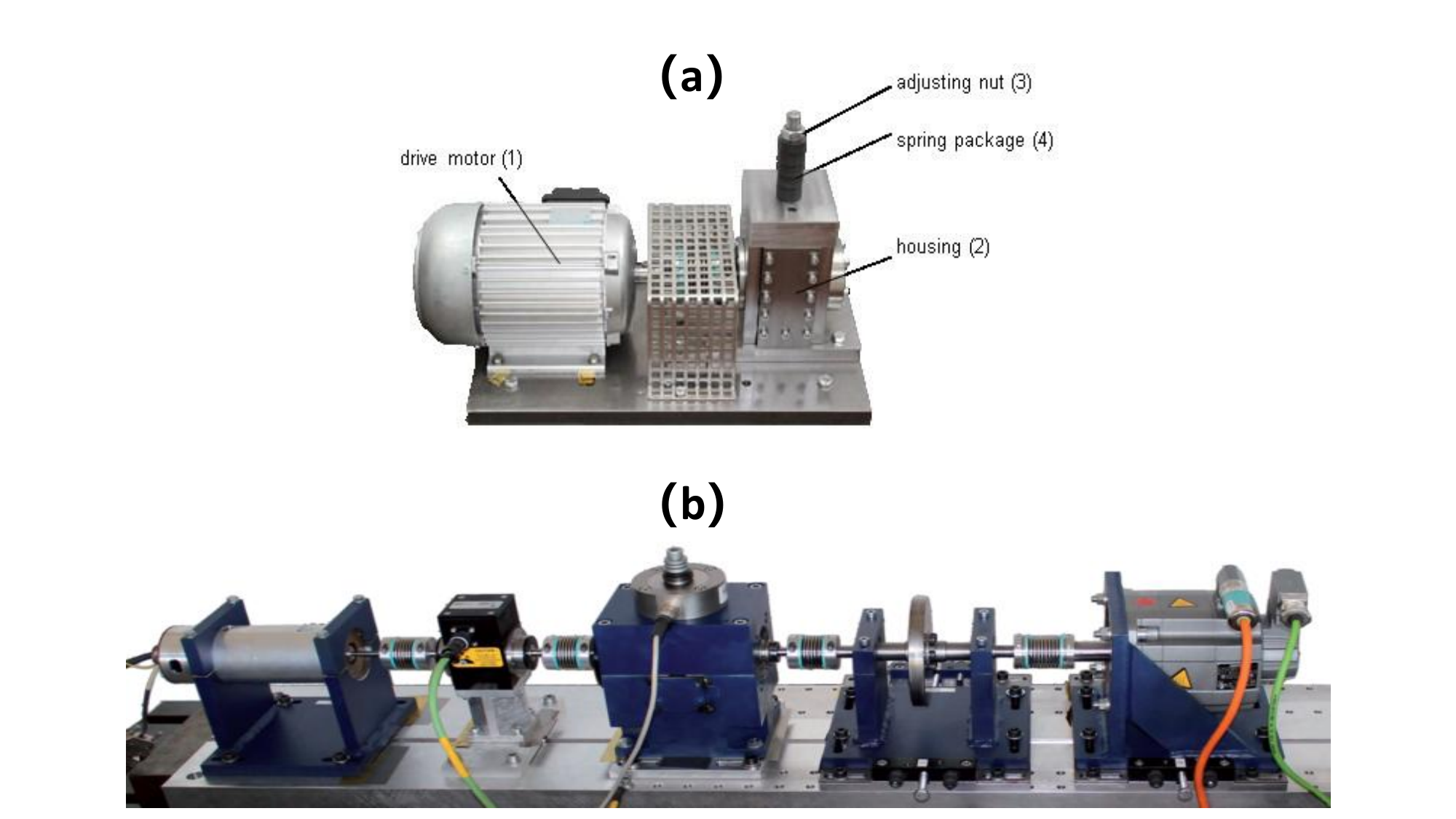}
	\caption{\label{fig12} The experimental machinery of KAt Research Data Center at Padrrborn University. (a) The setup for the accelerated lifetime test. (b) The test rig to collect the vibration signals.}
\end{figure}

In this section, the fault bearing dataset from the KAt Research Data Center at Paderborn University is utilized for diagnosis. The goal is to classify the fault types of bearings based on features extracted using LZC-based methods. For our experiments, real damaged bearings from an accelerated lifetime test are employed. The setup for the accelerated lifetime test is illustrated in Fig.~\ref{fig12}(a). This rig consists of a bearing housing and an electric motor that powers the shaft, which supports four test bearings of type 6203. The bearings are subjected to radial force generated by a spring-screw mechanism.

The vibration signals are collected from the test rig depicted in Fig.~\ref{fig12}(b), which includes a drive motor, a torque measurement shaft, a bearing module, a flywheel, and a load motor. Data is selected from a basic configuration of the operational parameters. The test rig operates at 1,500 rpm with a load torque of 0.7 Nm and applies a radial force of 1,000 N to the bearing. The sampling frequency is set at 64 kHz. 

In our experiment, nine types of fault bearings are selected for diagnosis, with detailed information provided in Table.~\ref{tab2}. In Table.~\ref{tab2}, "OR" denotes outer ring fault, and "IR" denotes inner ring fault. The notations "S," "R," and "M" represent single damage, repetitive damage, and multiple damage, respectively. For each bearing, 100 random samples of vibration signals, each with a length of $N=5000$, are collected for classification. Four methods (LZC, PLZC, DELZC, BT-DELZC) are used to extract features and perform comparisons. The parameters for PLZC, DELZC, and BT-DELZC are set to $m=3$, $c=5$ and $\tau=2$.

\begin{table}
	\centering
	\caption{The detailed information of the 9 fault bearings conditions in KAt Data Center at Paderborn University.}
	\resizebox{0.8\linewidth}{!}{\begin{tabular}{ccccccc}
			\toprule
			\textbf{Bearing code} & \textbf{Damage} & \textbf{Bearing element} & \textbf{Combination} 
			& \textbf{Arrangement} & \textbf{Damage extent} & \textbf{Characteristic} \\
			\midrule
			KA04 & fatigue: pitting & OR & S & no repetition & 1 & single point \\
			KA16 & fatigue: pitting & OR & R & random & 2 & single point \\
			KA30 & Plastic deform & OR & R & random & 1 & distributed \\
			KB23 & fatigue: pitting & IR(+OR) & M & random & 2 & single point \\
			KB24 & fatigue: pitting & IR(+OR) & M & no repetition & 3 & distributed \\
			KB27 & Plastic deform & OR+IR & M & random & 1 & distributed \\
			KI04 & fatigue: pitting & IR & M & no repetition & 1 & single point \\
			KI14 & fatigue: pitting & IR & M & no repetition & 1 & single point \\
			KI17 & fatigue: pitting & IR & R & random & 1 & single point \\
			
			\bottomrule
			\label{tab2}
	\end{tabular}}
\end{table}

\begin{figure*}
	\centering
	\includegraphics[width=6.5in]{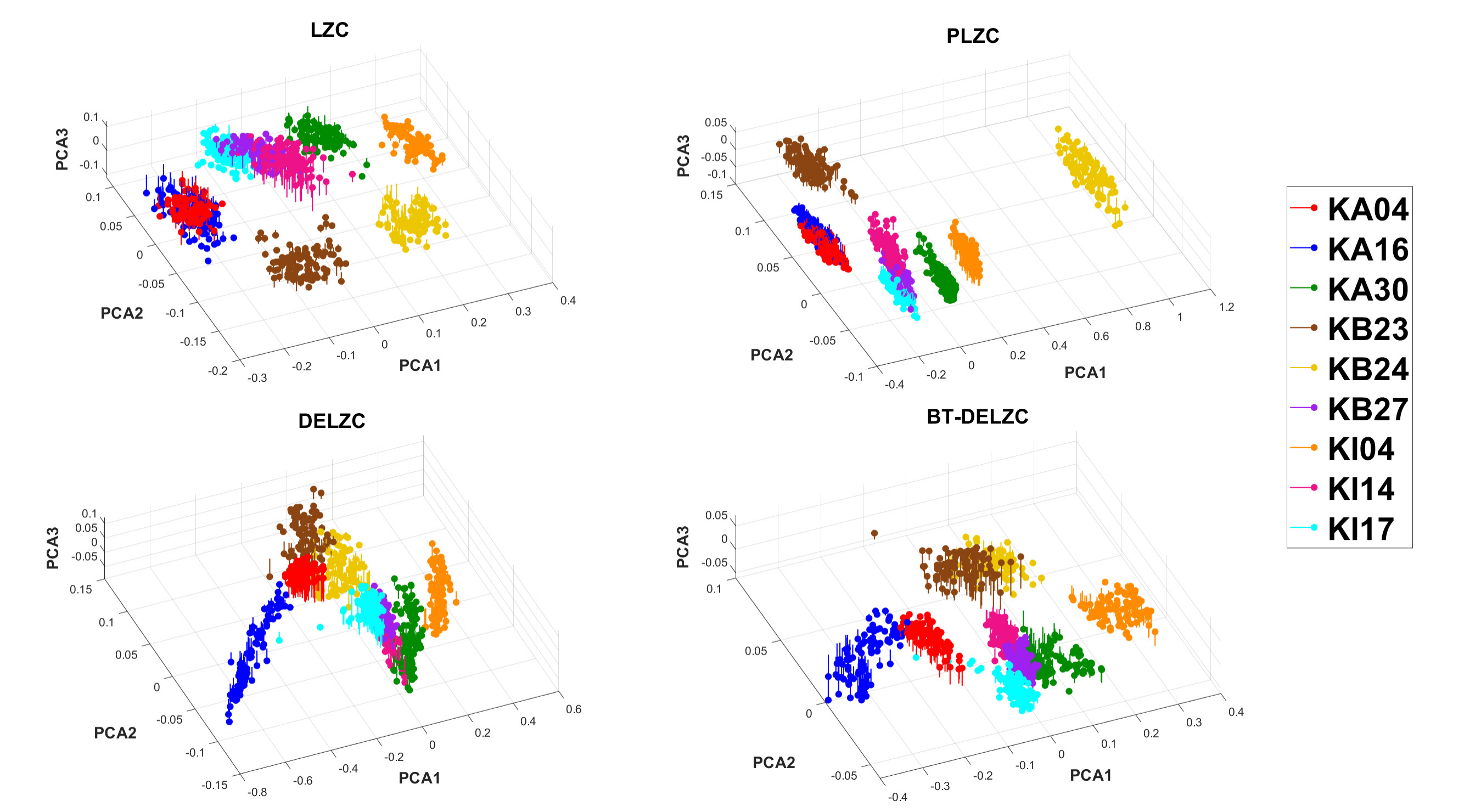}
	\caption{\label{fig13} The needle charts of first three features extracted by the PCA algorithm from vibration signals of different fault bearings from KAt Data Centers at Paderborn University for four methods (LZC, PLZC, DELZC, BT-DELZC)}
\end{figure*}

Fig.~\ref{fig13} illustrates the first three features extracted by the PCA algorithm from vibration signals of different fault bearings, following the procedures outlined in Section~\ref{sec4a}. The features are visualized using a needle chart, where the direction of the needle tip indicates positive or negative values. The results reveal that the PLZC and DELZC methods struggle to distinguish between the KB27, KI14, and KI17 bearing types. Similarly, LZC has difficulty differentiating between KB27 and KI14. Additionally, LZC and PLZC methods show overlapping features for KA04 and KA16, making these types hard to distinguish. In contrast, the proposed BT-DELZC method stands out by effectively differentiating all nine bearing types to a greater extent, as demonstrated in the needle chart. 

\begin{figure}
	\centering
	\includegraphics[width=3.5in]{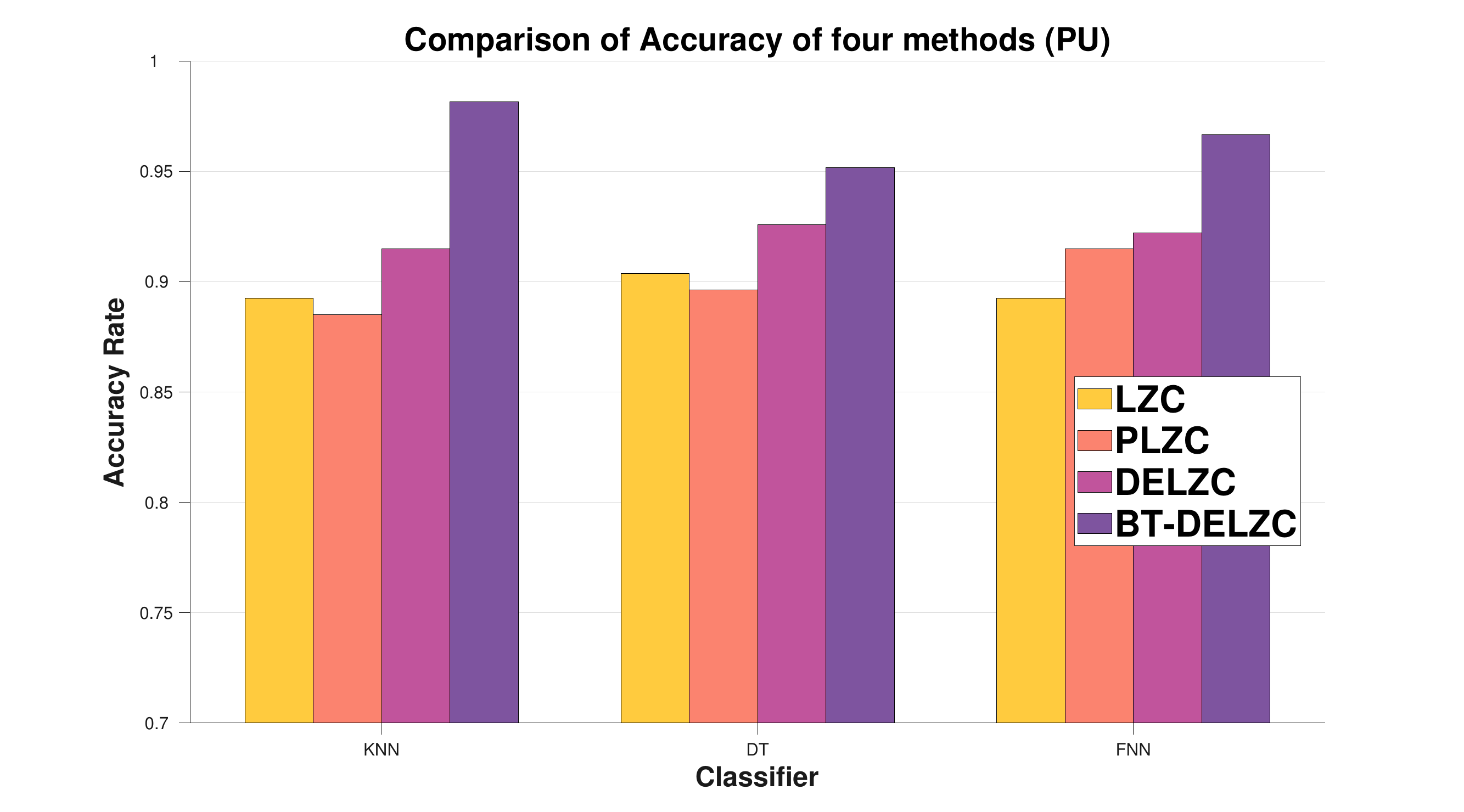}
	\caption{\label{fig14} The accuracy rate results of four methods (LZC, PLZC, DELZC, BT-DELZC) under three classifiers (KNN, DT, FNN)}
\end{figure}

Next, machine learning classifiers—specifically, K-Nearest Neighbor (KNN), Decision Tree (DT), and Feedforward Neural Network (FNN) models—are employed for classification. The 3-dimensional features for each bearing type are split into a training set (70 percent) and a test set (30 percent). The training sets are combined and fed into the classifiers, along with their bearing type labels, to train the models (the classifiers are provided by the Statistics and Machine Learning Toolbox in MATLAB 2021a). The test sets are then used for classification to determine the accuracy rates, which are presented in Fig.~\ref{fig14} and Table.~\ref{tab3}. The results indicate that the proposed BT-DELZC method achieves the highest accuracy rates among the four methods.

\begin{table}
	\centering
	\caption{The accuracy rate results of PU bearing datasets fault diagnosis for four methods (LZC, PLZC, DELZC, BT-DELZC) under three classifiers.}
	\resizebox{0.6\linewidth}{!}{\begin{tabular}{cc|rrrr}
			\toprule
			\multicolumn{2}{c}{\multirow{2}{*}{\textbf{Classifier}}} & 
			\multicolumn{4}{c}{\textbf{LZC method}} \\
			\cmidrule{3-6} \multicolumn{2}{c}{} &
			\multicolumn{1}{c}{LZC} & \multicolumn{1}{c}{PLZC} &
			\multicolumn{1}{c}{DELZC} & \multicolumn{1}{c}{BT-DELZC} 
			\\
			
			\midrule
			
			\multicolumn{2}{c}{\textbf{KNN}} &  \multicolumn{1}{c}{89.26\%} & \multicolumn{1}{c}{88.52\%} & \multicolumn{1}{c}{91.48\%} & \multicolumn{1}{c}{\textbf{98.15\%}}   \\
			\midrule 
			
			\multicolumn{2}{c}{\textbf{DT}} &  \multicolumn{1}{c}{90.37\%} & \multicolumn{1}{c}{89.63\%} & \multicolumn{1}{c}{92.59\%} & \multicolumn{1}{c}{\textbf{95.19\%}}   \\
			\midrule 
			
			\multicolumn{2}{c}{\textbf{FNN}} &  \multicolumn{1}{c}{89.26\%} & \multicolumn{1}{c}{91.48\%} & \multicolumn{1}{c}{92.22\%} & \multicolumn{1}{c}{\textbf{96.67\%}}   \\

			\bottomrule
			\label{tab3}
	\end{tabular}}
\end{table}

\begin{figure}
	\centering
	\includegraphics[width=4.5in]{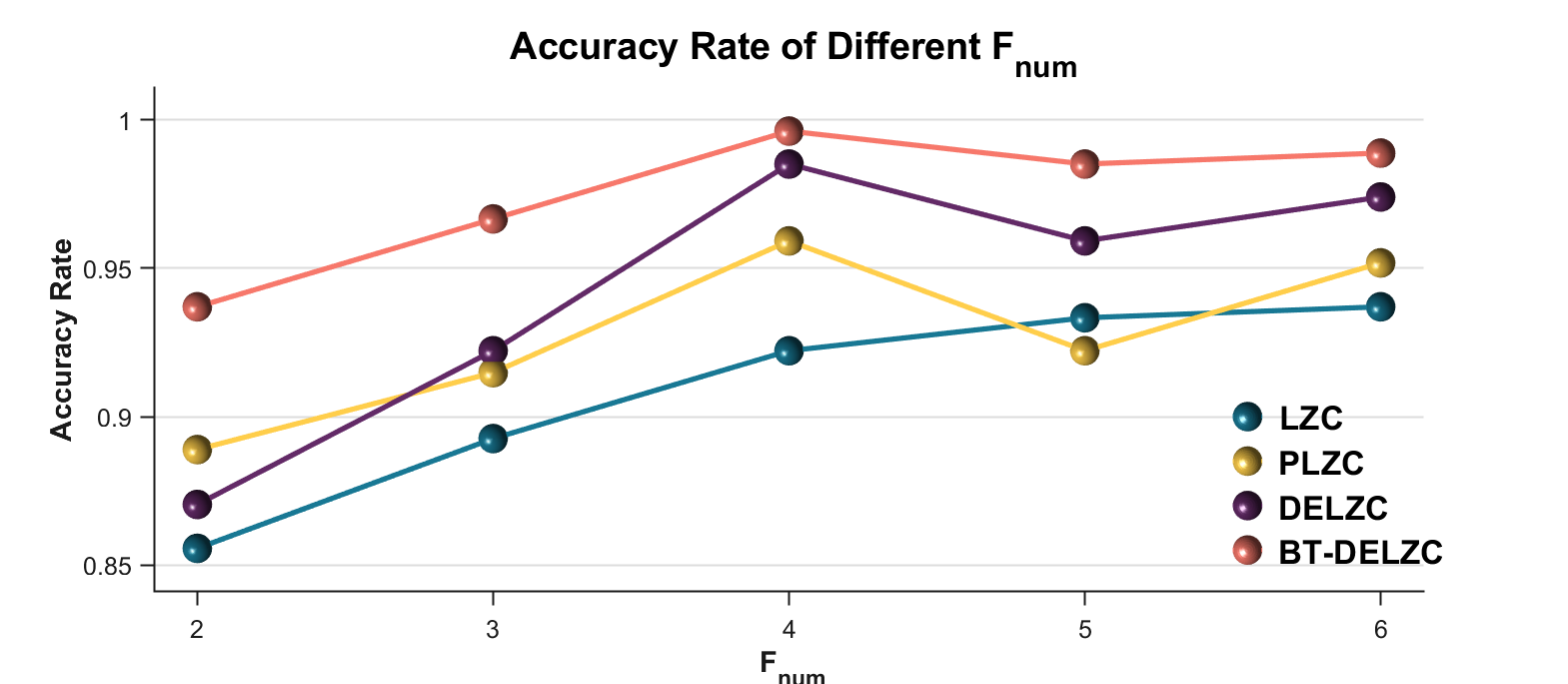}
	\caption{\label{fig15} The accuracy rate results of PU bearing dataset for four methods (LZC, PLZC, DELZC, BT-DELZC) by FNN classifier under the different $F_{num}$}
\end{figure}

Additionally, the Feedforward Neural Network (FNN) models are used to compute accuracy rates with varying numbers of principal components $F_{num}$ extracted via PCA. The range of $F_{num}$ tested is from 2 to 6. The accuracy results for these different $F_{num}$ selections are illustrated in Fig.~\ref{fig15} and detailed in Table.~\ref{tab4}. The findings reveal that the proposed BT-DELZC method consistently outperforms the other three LZC-based metrics across all tested values of $F_{num}$. Overall, these experimental results underscore the superior performance of BT-DELZC in feature extraction from nonlinear signals and time series data.

\begin{table}
	\centering
	\caption{The accuracy rate results with varying numbers of principal components $F_{num}$ for four methods (LZC, PLZC, DELZC, BT-DELZC) by FNN classifiers. }
	\resizebox{0.6\linewidth}{!}{\begin{tabular}{cc|rrrr}
			\toprule
			\multicolumn{2}{c}{\multirow{2}{*}{\textbf{$F_{num}$}}} & 
			\multicolumn{4}{c}{\textbf{LZC method}} \\
			\cmidrule{3-6} \multicolumn{2}{c}{} &
			\multicolumn{1}{c}{LZC} & \multicolumn{1}{c}{PLZC} &
			\multicolumn{1}{c}{DELZC} & \multicolumn{1}{c}{BT-DELZC}           \\
			
			\midrule
			
			\multicolumn{2}{c}{\textbf{2}} &  \multicolumn{1}{c}{85.56\%} & \multicolumn{1}{c}{88.89\%} & \multicolumn{1}{c}{87.04\%} & \multicolumn{1}{c}{\textbf{93.70\%}}   \\
			
			\midrule 
			
			\multicolumn{2}{c}{\textbf{3}} &  \multicolumn{1}{c}{89.26\%} & \multicolumn{1}{c}{91.48\%} & \multicolumn{1}{c}{92.22\%} & \multicolumn{1}{c}{\textbf{96.67\%}}   \\
			
			\midrule 
			
			\multicolumn{2}{c}{\textbf{4}} &  \multicolumn{1}{c}{92.22\%} & \multicolumn{1}{c}{95.93\%} & \multicolumn{1}{c}{98.52\%} & \multicolumn{1}{c}{\textbf{99.63\%}}   \\
			
			\midrule 
			
			\multicolumn{2}{c}{\textbf{5}} &  \multicolumn{1}{c}{93.33\%} & \multicolumn{1}{c}{92.22\%} & \multicolumn{1}{c}{95.93\%} & \multicolumn{1}{c}{\textbf{98.52\%}}   \\
			
			\midrule 
			
			\multicolumn{2}{c}{\textbf{6}} &  \multicolumn{1}{c}{93.70\%} & \multicolumn{1}{c}{95.19\%} & \multicolumn{1}{c}{97.41\%} & \multicolumn{1}{c}{\textbf{98.89\%}}   \\

			\bottomrule
			\label{tab4}
	\end{tabular}}
\end{table}

\subsection{Case Western Reserve University Fault Bearing Diagnosis}
\label{sec4c}
\begin{figure}
	\centering
	\includegraphics[width=3in]{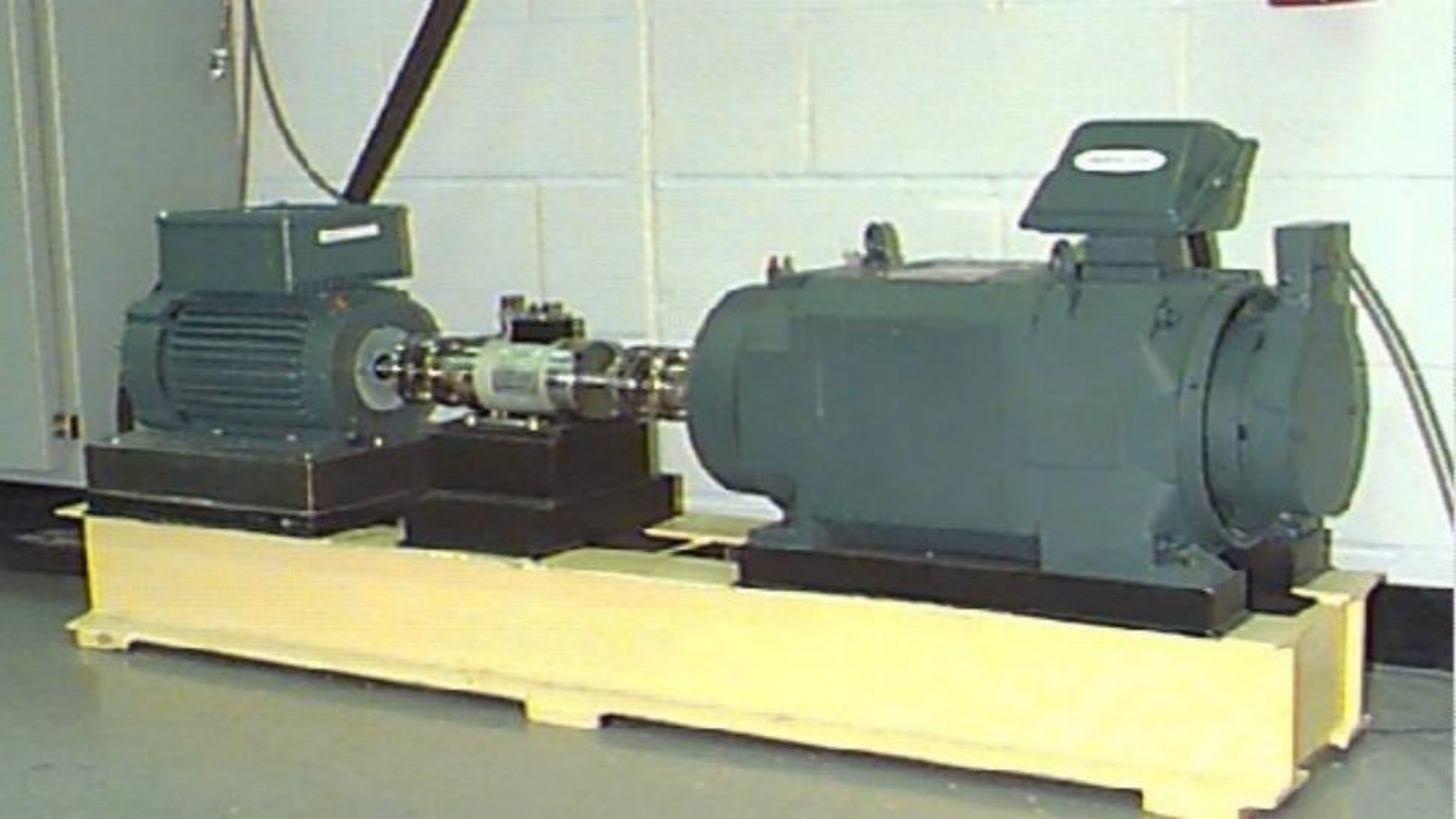}
	\caption{\label{fig16} The bearing test stand of Case Western Reserve University. }
\end{figure}

The second experiment evaluates the proposed BT-DELZC method using the fault bearing dataset from Case Western Reserve University. As shown in Fig.~\ref{fig16}, the test setup includes a 2 hp motor, a torque transducer, a dynamometer, and control electronics (not depicted). Vibration signals are recorded from accelerometers mounted on the fan end, with a sampling frequency of 12 kHz and a motor speed of 1797 revolutions per minute (r/min).

The experiment examines 12 single-point fault types, varying in location and damage degree. Detailed information on the bearing codes and fault types is provided in Table.~\ref{tab5}. For each fault type, 100 signals, each with a length of $N=5000$, are randomly sampled from the original data. Classification is performed as outlined in Section~\ref{sec4a}, and a comparison is made using four methods: LZC, PLZC, DELZC, and BT-DELZC. The parameters for PLZC, DELZC, and BT-DELZC are set to $m=4$, $c=4$ and $\tau=2$.

\begin{table}
	\centering
	\caption{The detailed information of the 12 fault-bearing conditions in Case Western Reserve University.}
	\resizebox{0.6\linewidth}{!}{\begin{tabular}{cccc}
			\toprule
			\textbf{Label} & \textbf{Bearing code} & \textbf{Fault component} & \textbf{Fault diameter}  \\
			\midrule
			1 & IR-7 & Inner Race & 0.007 inches   \\
			2 & IR-14 & Inner Race & 0.014 inches   \\
			3 & IR-21 & Inner Race & 0.021 inches  \\
			4 & BA-7 & Ball & 0.007 inches   \\
			5 & BA-14 & Ball & 0.014 inches  \\
			6 & BA-21 & Ball & 0.021 inches  \\
			7 & OR6-7 & Outer Race (6 o'clock) & 0.007 inches   \\
			8 & OR6-14 & Outer Race (6 o'clock) & 0.014 inches  \\
			9 & OR6-21 & Outer Race (6 o'clock) & 0.021 inches   \\
			10 & OR3-7 & Outer Race (3 o'clock)& 0.007 inches  \\
			11 & OR3-14 & Outer Race (3 o'clock)& 0.014 inches  \\
			12 & OR12-7 & Outer Race (12 o'clock)& 0.007 inches  \\
			
			\bottomrule
			\label{tab5}
	\end{tabular}}
\end{table}

\begin{figure*}
	\centering
	\includegraphics[width=6.5in]{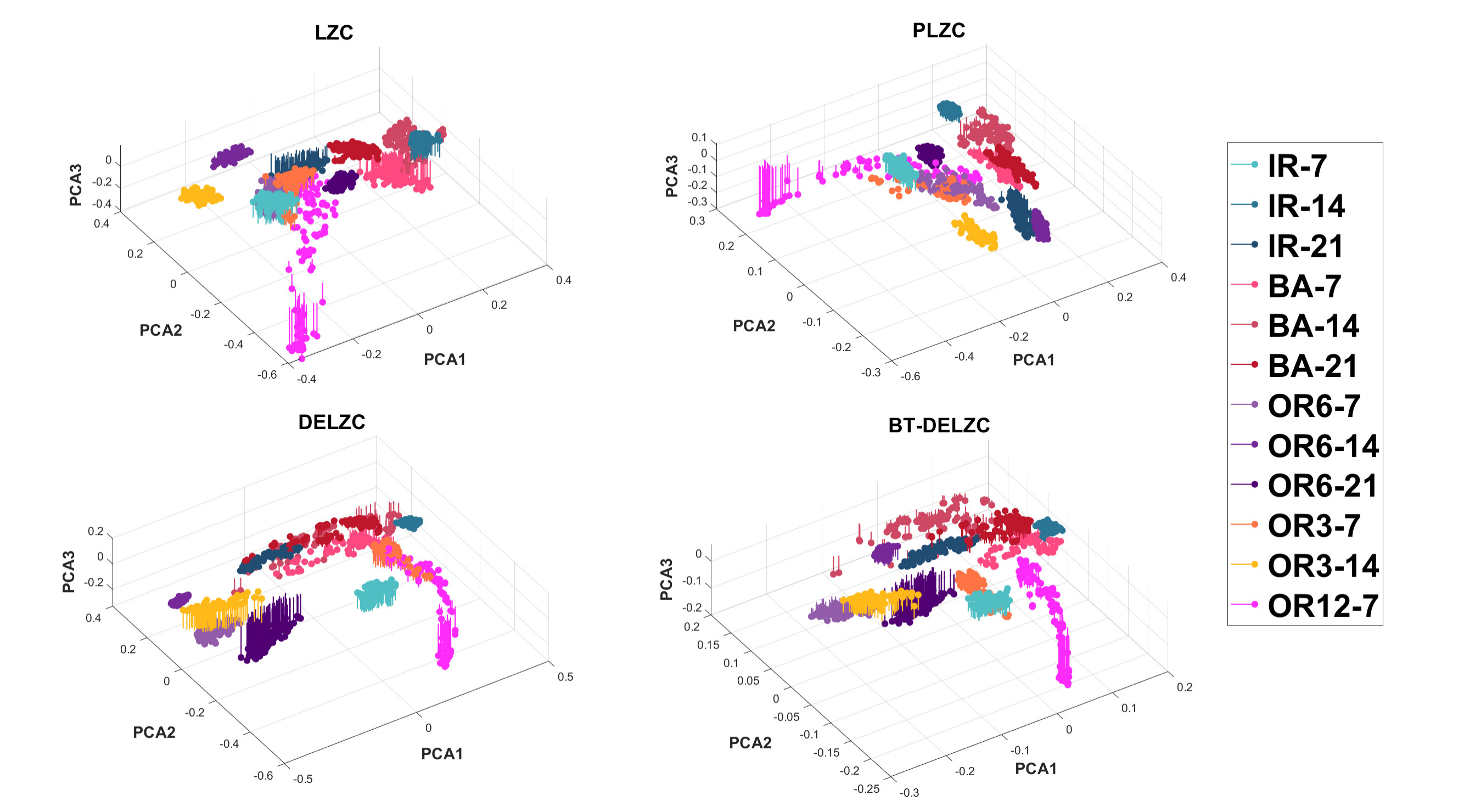}
	\caption{\label{fig17} The needle chart of first three features extracted by the PCA algorithm from fan end vibration signals of different fault bearings from Case Western Reserve University for four methods (LZC, PLZC, DELZC, BT-DELZC).}
\end{figure*}

The results of the 3-dimensional PCA features from the samples are presented in Fig.~\ref{fig17} using needle charts. It is evident that LZC, PLZC, and DELZC struggle to distinguish all fault bearing types, with some samples overlapping (e.g., OR6-7 and OR3-7 in LZC and PLZC, OR3-7 and OR12-7 in DELZC). In contrast, BT-DELZC demonstrates the most effective capability in classifying fault bearing types among the four methods.

\begin{figure}
	\centering
	\includegraphics[width=3.5in]{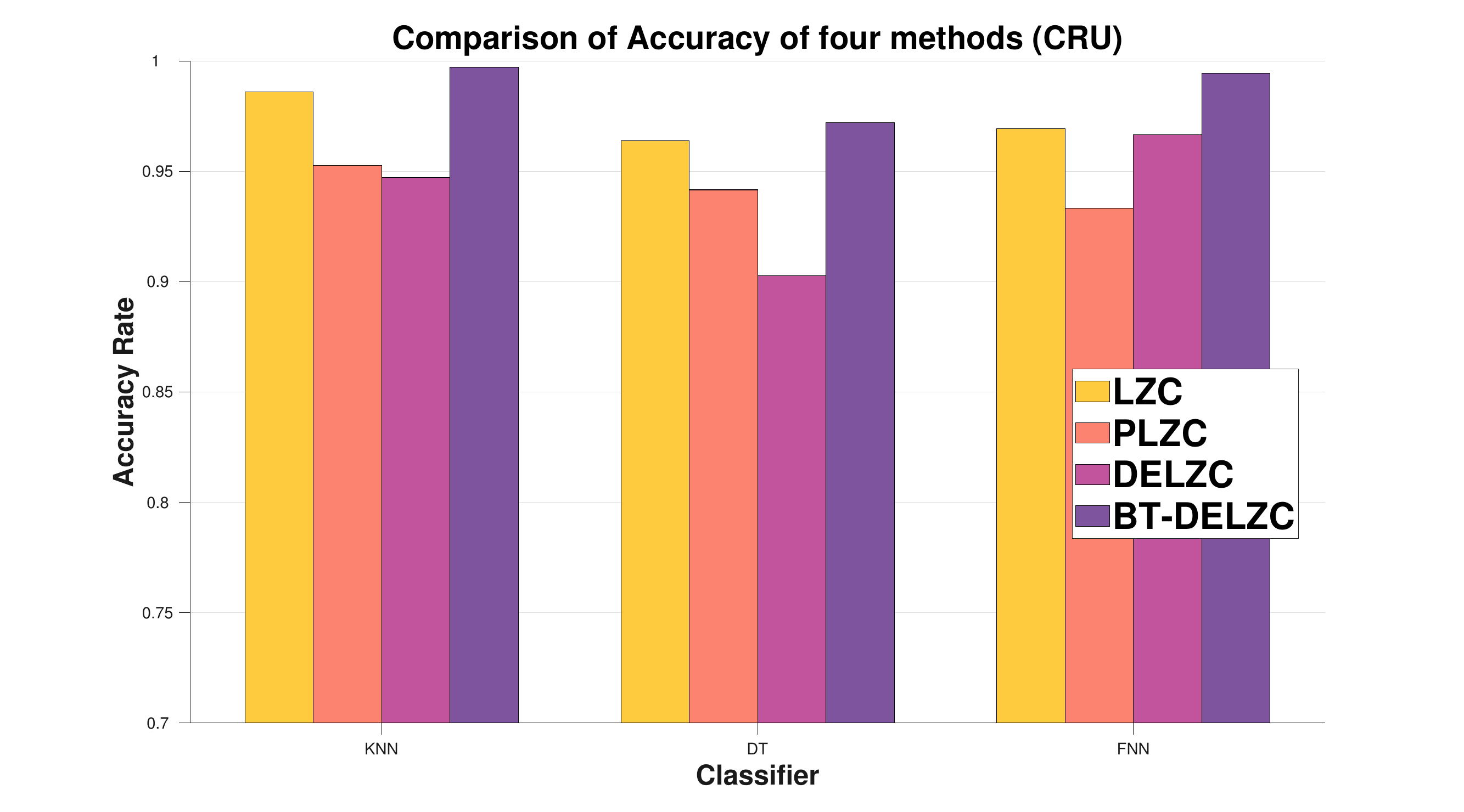}
	\caption{\label{fig18} The accuracy rate results of CWRU bearing datasets for four methods (LZC, PLZC, DELZC, BT-DELZC) under three classifiers (KNN, DT, FNN)}
\end{figure}

\begin{figure}
	\centering
	\includegraphics[width=5in]{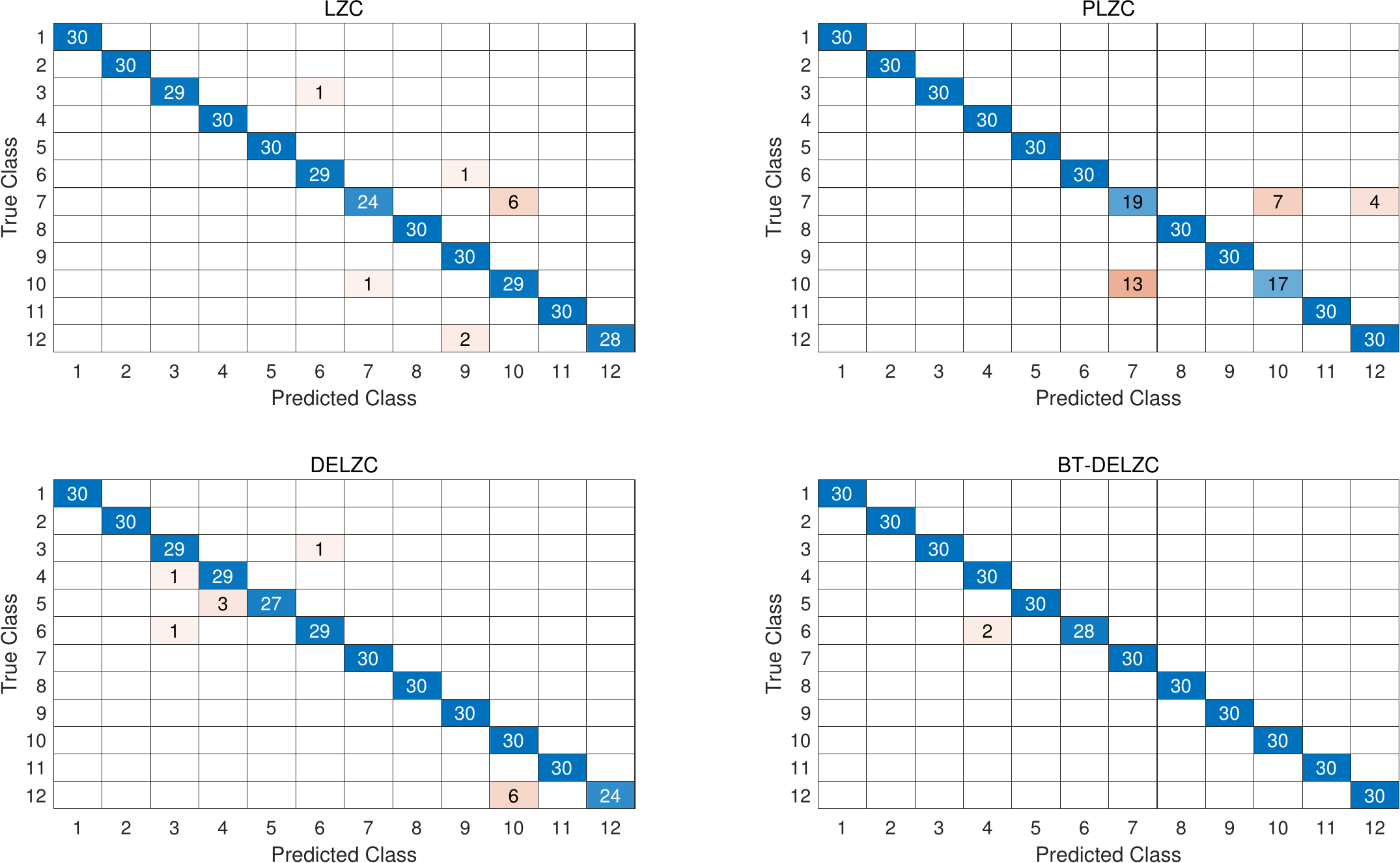}
	\caption{\label{fig19} The confusion matrices for four methods (LZC, PLZC, DELZC, BT-DELZC) of CWRU bearing dataset under FNN classifier.}
\end{figure}

Finally, the accuracy rates for the four methods, evaluated using three classifiers (KNN, DT, and FNN), are shown in Fig.~\ref{fig18} and detailed in Table.~\ref{tab6}. The confusion matrices for the FNN classifier are illustrated in Fig.~\ref{fig19}. The results indicate that BT-DELZC consistently outperforms the other methods in distinguishing fault bearing types. This strongly demonstrates the effectiveness of the BT-DELZC method in recognizing characteristics from nonlinear time series data.  

\begin{table}
	\centering
	\caption{The accuracy rate results of CWRU bearing datasets fault diagnosis for four methods (LZC, PLZC, DELZC, DETLZC) under three classifiers.}
	\resizebox{0.6\linewidth}{!}{\begin{tabular}{cc|rrrr}
			\toprule
			\multicolumn{2}{c}{\multirow{2}{*}{\textbf{Classifier}}} & 
			\multicolumn{4}{c}{\textbf{LZC method}} \\
			\cmidrule{3-6} \multicolumn{2}{c}{} &
			\multicolumn{1}{c}{LZC} & \multicolumn{1}{c}{PLZC} &
			\multicolumn{1}{c}{DELZC} & \multicolumn{1}{c}{BT-DELZC} 
			\\
			
			\midrule
			
			\multicolumn{2}{c}{\textbf{KNN}} &  \multicolumn{1}{c}{98.61\%} & \multicolumn{1}{c}{95.28\%} & \multicolumn{1}{c}{94.72\%} & \multicolumn{1}{c}{\textbf{99.72\%}}   \\
			\midrule 
			
			\multicolumn{2}{c}{\textbf{DT}} &  \multicolumn{1}{c}{96.39\%} & \multicolumn{1}{c}{94.17\%} & \multicolumn{1}{c}{90.28\%} & \multicolumn{1}{c}{\textbf{97.22\%}}   \\
			\midrule 
			
			\multicolumn{2}{c}{\textbf{FNN}} &  \multicolumn{1}{c}{96.94\%} & \multicolumn{1}{c}{93.33\%} & \multicolumn{1}{c}{96.67\%} & \multicolumn{1}{c}{\textbf{99.44\%}}   \\

			\bottomrule
			\label{tab6}
	\end{tabular}}
\end{table}

\section{Discussion}
\label{dis} 
This paper introduces a novel approach to nonlinear time series analysis based on the LZC algorithm, marking the first attempt to integrate LZC-based metrics with time series network methods.

Recent developments have seen time series network methods—such as transition networks, visibility graphs, and recurrence plots—combined with entropy algorithms. These combinations, as referenced in the introduction, have led to effective new measures for characterizing nonlinear time series. Since LZC-based methods also involve a symbolization preprocessing step, it is natural to extend time series network methods to include LZC-based metrics.

Among time series network methods, transition networks provide a direct way to utilize embedding vectors by focusing on transition patterns within symbolic sequences. This approach captures dynamic behaviors in nonlinear time series more effectively than traditional methods, enhancing sequence recognition and differentiation. Consequently, the BT-DELZC algorithm excels in analyzing gradually changing time series, such as chirp signals and mixed processes, as discussed in Section~\ref{sec3}.

In our design, the dispersion pattern of embedding vectors is chosen for its ability to account for amplitude values in time series. This method has demonstrated superior performance in previous nonlinear analysis research. The transition framework construction is also explored. Self-loops are omitted to better highlight dynamic structures, although their inclusion remains a valuable area for future research, as they can dominate chaotic systems. Additionally, our algorithm focuses on link sequences between dispersion patterns and calculates their weighted sums. Future research could explore constructing symbolic sequences by merging all transition patterns, offering an alternative framework worth investigating.

\section{Conclusion}
\label{con}
In this paper, we introduce a novel method called Bidirectional Transition Dispersion Entropy-based Lempel-Ziv Complexity (BT-DELZC) designed to better characterize the structures of nonlinear time series. The BT-DELZC method integrates the DELZC approach with a transition network framework, constructing symbolic sequences from a weighted and link-based perspective to capture dynamic structures between adjacent embedding vectors. Additionally, hierarchical decomposition is employed to extract more features from the original time series across different frequency components.

In the simulated data experiments, we first explore parameter selection by varying signal lengths. We then demonstrate the robustness of BT-DELZC by calculating standard deviations across different lengths of white noise and assessing its ability to distinguish between fully chaotic logistic maps and white noise under varying SNR. The chirp signal and mixed process are used to validate the effectiveness of BT-DELZC in capturing dynamic characteristics of time series. The results indicate that BT-DELZC consistently outperforms previous methods such as LZC, PLZC, and DELZC.

Finally, we apply BT-DELZC to two real-world fault bearing datasets to evaluate its classification performance for different signal types. By utilizing hierarchical decomposition, PCA, and machine learning models to extract LZC-based features from signal samples, we find that BT-DELZC achieves the highest accuracy rates in all scenarios. This underscores the superiority of BT-DELZC in distinguishing various fault bearings.

In summary, BT-DELZC demonstrates significant improvements over previous methods (LZC, PLZC, and DELZC) in both simulated data and real-world bearing data experiments. Its accuracy and effectiveness in feature extraction from nonlinear time series are well-established, making it a valuable tool for industrial engineering and other fields. BT-DELZC offers a fresh perspective for analyzing signal complexity and holds promise for future research in nonlinear systems.

\section*{CRediT authorship contribution statement}
\textbf{Runze jiang:} Conceptualization; Data Curation; Formal analysis; Methodology; Software; Validation; Visualization; Writing/Original draft preparation; Writing/Review \& Editing.  

\textbf{Pengjian Shang:} Formal analysis; Funding Acquisition; Methodology; Visualization; Writing/Review \& Editing.

\section*{Declaration of Competing Interest}
The authors declare that they have no known competing financial 
interests or personal relationships that could have appeared to influence the work reported in this paper.

\section*{Data availability}
Data will be made available on request.

\section*{Acknowledgment}

The financial supports from the Fundamental Research Funds for the Central Universities, China (2024YJS164), and the National Natural Science Foundation of China (62171018).








\printcredits

\bibliographystyle{cas-model2-names}

\bibliography{ref.bib}

\begin{thebibliography}{39}
\expandafter\ifx\csname natexlab\endcsname\relax\def\natexlab#1{#1}\fi
\providecommand{\url}[1]{\texttt{#1}}
\providecommand{\href}[2]{#2}
\providecommand{\path}[1]{#1}
\providecommand{\DOIprefix}{doi:}
\providecommand{\ArXivprefix}{arXiv:}
\providecommand{\URLprefix}{URL: }
\providecommand{\Pubmedprefix}{pmid:}
\providecommand{\doi}[1]{\href{http://dx.doi.org/#1}{\path{#1}}}
\providecommand{\Pubmed}[1]{\href{pmid:#1}{\path{#1}}}
\providecommand{\bibinfo}[2]{#2}
\ifx\xfnm\relax \def\xfnm[#1]{\unskip,\space#1}\fi
\bibitem[{Aboy et~al.(2006)Aboy, Hornero, Ab{\'a}solo and {\'A}lvarez}]{LZC1}
\bibinfo{author}{Aboy, M.}, \bibinfo{author}{Hornero, R.},
  \bibinfo{author}{Ab{\'a}solo, D.}, \bibinfo{author}{{\'A}lvarez, D.},
  \bibinfo{year}{2006}.
\newblock \bibinfo{title}{Interpretation of the lempel-ziv complexity measure
  in the context of biomedical signal analysis}.
\newblock \bibinfo{journal}{IEEE transactions on biomedical engineering}
  \bibinfo{volume}{53}, \bibinfo{pages}{2282--2288}.
\bibitem[{Azami et~al.(2017)Azami, Fern{\'a}ndez and Escudero}]{chirp}
\bibinfo{author}{Azami, H.}, \bibinfo{author}{Fern{\'a}ndez, A.},
  \bibinfo{author}{Escudero, J.}, \bibinfo{year}{2017}.
\newblock \bibinfo{title}{Refined multiscale fuzzy entropy based on standard
  deviation for biomedical signal analysis}.
\newblock \bibinfo{journal}{Medical \& biological engineering \& computing}
  \bibinfo{volume}{55}, \bibinfo{pages}{2037--2052}.
\bibitem[{Bai et~al.(2015)Bai, Liang and Li}]{PLZC}
\bibinfo{author}{Bai, Y.}, \bibinfo{author}{Liang, Z.}, \bibinfo{author}{Li,
  X.}, \bibinfo{year}{2015}.
\newblock \bibinfo{title}{A permutation lempel-ziv complexity measure for eeg
  analysis}.
\newblock \bibinfo{journal}{Biomedical Signal Processing and Control}
  \bibinfo{volume}{19}, \bibinfo{pages}{102--114}.
\bibitem[{Bandt and Pompe(2002)}]{PE}
\bibinfo{author}{Bandt, C.}, \bibinfo{author}{Pompe, B.}, \bibinfo{year}{2002}.
\newblock \bibinfo{title}{Permutation entropy: a natural complexity measure for
  time series}.
\newblock \bibinfo{journal}{Physical review letters} \bibinfo{volume}{88},
  \bibinfo{pages}{174102}.
\bibitem[{Borowska(2021)}]{SLZC}
\bibinfo{author}{Borowska, M.}, \bibinfo{year}{2021}.
\newblock \bibinfo{title}{Multiscale permutation lempel--ziv complexity measure
  for biomedical signal analysis: Interpretation and application to focal eeg
  signals}.
\newblock \bibinfo{journal}{Entropy} \bibinfo{volume}{23},
  \bibinfo{pages}{832}.
\bibitem[{Costa et~al.(2005)Costa, Goldberger and Peng}]{multiscale}
\bibinfo{author}{Costa, M.}, \bibinfo{author}{Goldberger, A.L.},
  \bibinfo{author}{Peng, C.K.}, \bibinfo{year}{2005}.
\newblock \bibinfo{title}{Multiscale entropy analysis of biological signals}.
\newblock \bibinfo{journal}{Physical Review E—Statistical, Nonlinear, and
  Soft Matter Physics} \bibinfo{volume}{71}, \bibinfo{pages}{021906}.
\bibitem[{Ding et~al.(2023)Ding, Ji, Li, Wang, Noman and Feng}]{LDE}
\bibinfo{author}{Ding, L.}, \bibinfo{author}{Ji, J.}, \bibinfo{author}{Li, Y.},
  \bibinfo{author}{Wang, S.}, \bibinfo{author}{Noman, K.},
  \bibinfo{author}{Feng, K.}, \bibinfo{year}{2023}.
\newblock \bibinfo{title}{A novel weak feature extraction method for rotating
  machinery: link dispersion entropy}.
\newblock \bibinfo{journal}{IEEE Transactions on Instrumentation and
  Measurement} .
\bibitem[{Ferrario et~al.(2005)Ferrario, Signorini, Magenes and Cerutti}]{mix}
\bibinfo{author}{Ferrario, M.}, \bibinfo{author}{Signorini, M.G.},
  \bibinfo{author}{Magenes, G.}, \bibinfo{author}{Cerutti, S.},
  \bibinfo{year}{2005}.
\newblock \bibinfo{title}{Comparison of entropy-based regularity estimators:
  application to the fetal heart rate signal for the identification of fetal
  distress}.
\newblock \bibinfo{journal}{IEEE Transactions on Biomedical Engineering}
  \bibinfo{volume}{53}, \bibinfo{pages}{119--125}.
\bibitem[{Ge and Lin(2023)}]{nd6}
\bibinfo{author}{Ge, X.}, \bibinfo{author}{Lin, A.}, \bibinfo{year}{2023}.
\newblock \bibinfo{title}{Quantifying the direct and indirect interactions for
  eeg signals by using detrended permutation mutual information}.
\newblock \bibinfo{journal}{Chaos, Solitons \& Fractals} \bibinfo{volume}{176},
  \bibinfo{pages}{114155}.
\bibitem[{Guo et~al.(2022)Guo, Ma, Zou, Gui and Li}]{bear4}
\bibinfo{author}{Guo, J.}, \bibinfo{author}{Ma, B.}, \bibinfo{author}{Zou, T.},
  \bibinfo{author}{Gui, L.}, \bibinfo{author}{Li, Y.}, \bibinfo{year}{2022}.
\newblock \bibinfo{title}{Composite multiscale transition permutation
  entropy-based fault diagnosis of bearings}.
\newblock \bibinfo{journal}{Sensors} \bibinfo{volume}{22},
  \bibinfo{pages}{7809}.
\bibitem[{Han et~al.(2021)Han, Wu, Wang and Liu}]{bear1}
\bibinfo{author}{Han, M.}, \bibinfo{author}{Wu, Y.}, \bibinfo{author}{Wang,
  Y.}, \bibinfo{author}{Liu, W.}, \bibinfo{year}{2021}.
\newblock \bibinfo{title}{Roller bearing fault diagnosis based on lmd and
  multi-scale symbolic dynamic information entropy}.
\newblock \bibinfo{journal}{Journal of Mechanical Science and Technology}
  \bibinfo{volume}{35}, \bibinfo{pages}{1993--2005}.
\bibitem[{He et~al.(2020)He, Shang and Zhang}]{nd8}
\bibinfo{author}{He, J.}, \bibinfo{author}{Shang, P.}, \bibinfo{author}{Zhang,
  Y.}, \bibinfo{year}{2020}.
\newblock \bibinfo{title}{Global recurrence quantification analysis and its
  application in financial time series}.
\newblock \bibinfo{journal}{Nonlinear Dynamics} \bibinfo{volume}{100},
  \bibinfo{pages}{803--829}.
\newblock \DOIprefix\doi{https://doi.org/10.1007/s11071-020-05543-4}.
\bibitem[{Huang et~al.(2012)Huang, Shang and Zhao}]{nd7}
\bibinfo{author}{Huang, J.}, \bibinfo{author}{Shang, P.},
  \bibinfo{author}{Zhao, X.}, \bibinfo{year}{2012}.
\newblock \bibinfo{title}{Multifractal diffusion entropy analysis on stock
  volatility in financial markets}.
\newblock \bibinfo{journal}{Physica A: Statistical Mechanics and its
  Applications} \bibinfo{volume}{391}, \bibinfo{pages}{5739--5745}.
\newblock \URLprefix
  \url{https://www.sciencedirect.com/science/article/pii/S0378437112005432},
  \DOIprefix\doi{https://doi.org/10.1016/j.physa.2012.06.039}.
\bibitem[{Huang et~al.(2021)Huang, Sun, Donner, Zhang, Guan and Zou}]{Huang}
\bibinfo{author}{Huang, M.}, \bibinfo{author}{Sun, Z.},
  \bibinfo{author}{Donner, R.V.}, \bibinfo{author}{Zhang, J.},
  \bibinfo{author}{Guan, S.}, \bibinfo{author}{Zou, Y.}, \bibinfo{year}{2021}.
\newblock \bibinfo{title}{Characterizing dynamical transitions by statistical
  complexity measures based on ordinal pattern transition networks}.
\newblock \bibinfo{journal}{Chaos: An Interdisciplinary Journal of Nonlinear
  Science} \bibinfo{volume}{31}.
\bibitem[{Huang et~al.(1998)Huang, Shen, Long, Wu, Shih, Zheng, Yen, Tung and
  Liu}]{emd}
\bibinfo{author}{Huang, N.E.}, \bibinfo{author}{Shen, Z.},
  \bibinfo{author}{Long, S.R.}, \bibinfo{author}{Wu, M.C.},
  \bibinfo{author}{Shih, H.H.}, \bibinfo{author}{Zheng, Q.},
  \bibinfo{author}{Yen, N.C.}, \bibinfo{author}{Tung, C.C.},
  \bibinfo{author}{Liu, H.H.}, \bibinfo{year}{1998}.
\newblock \bibinfo{title}{The empirical mode decomposition and the hilbert
  spectrum for nonlinear and non-stationary time series analysis}.
\newblock \bibinfo{journal}{Proceedings of the Royal Society of London. Series
  A: mathematical, physical and engineering sciences} \bibinfo{volume}{454},
  \bibinfo{pages}{903--995}.
\bibitem[{Jiang and Shang(2024)}]{nd3}
\bibinfo{author}{Jiang, R.}, \bibinfo{author}{Shang, P.}, \bibinfo{year}{2024}.
\newblock \bibinfo{title}{Dispersion complexity--entropy curves: An effective
  method to characterize the structures of nonlinear time series}.
\newblock \bibinfo{journal}{Chaos: An Interdisciplinary Journal of Nonlinear
  Science} \bibinfo{volume}{34}.
\bibitem[{Jiang et~al.(2011)Jiang, Peng and Xu}]{hier}
\bibinfo{author}{Jiang, Y.}, \bibinfo{author}{Peng, C.K.}, \bibinfo{author}{Xu,
  Y.}, \bibinfo{year}{2011}.
\newblock \bibinfo{title}{Hierarchical entropy analysis for biological
  signals}.
\newblock \bibinfo{journal}{Journal of Computational and Applied Mathematics}
  \bibinfo{volume}{236}, \bibinfo{pages}{728--742}.
\bibitem[{Lempel and Ziv(1976)}]{LZC}
\bibinfo{author}{Lempel, A.}, \bibinfo{author}{Ziv, J.}, \bibinfo{year}{1976}.
\newblock \bibinfo{title}{On the complexity of finite sequences}.
\newblock \bibinfo{journal}{IEEE Transactions on information theory}
  \bibinfo{volume}{22}, \bibinfo{pages}{75--81}.
\bibitem[{Lessmeier et~al.(2016)Lessmeier, Kimotho, Zimmer and Sextro}]{PU}
\bibinfo{author}{Lessmeier, C.}, \bibinfo{author}{Kimotho, J.K.},
  \bibinfo{author}{Zimmer, D.}, \bibinfo{author}{Sextro, W.},
  \bibinfo{year}{2016}.
\newblock \bibinfo{title}{Condition monitoring of bearing damage in
  electromechanical drive systems by using motor current signals of electric
  motors: A benchmark data set for data-driven classification}, in:
  \bibinfo{booktitle}{PHM Society European Conference}.
\bibitem[{Li et~al.(2022a)Li, Liu and Yang}]{nd4}
\bibinfo{author}{Li, G.}, \bibinfo{author}{Liu, F.}, \bibinfo{author}{Yang,
  H.}, \bibinfo{year}{2022}a.
\newblock \bibinfo{title}{Research on feature extraction method of ship
  radiated noise with k-nearest neighbor mutual information variational mode
  decomposition, neural network estimation time entropy and self-organizing map
  neural network}.
\newblock \bibinfo{journal}{Measurement} \bibinfo{volume}{199},
  \bibinfo{pages}{111446}.
\bibitem[{Li et~al.(2022b)Li, Geng and Jiao}]{DELZC}
\bibinfo{author}{Li, Y.}, \bibinfo{author}{Geng, B.}, \bibinfo{author}{Jiao,
  S.}, \bibinfo{year}{2022}b.
\newblock \bibinfo{title}{Dispersion entropy-based lempel-ziv complexity: A new
  metric for signal analysis}.
\newblock \bibinfo{journal}{Chaos, Solitons \& Fractals} \bibinfo{volume}{161},
  \bibinfo{pages}{112400}.
\bibitem[{Li et~al.(2023)Li, Geng and Tang}]{DE1}
\bibinfo{author}{Li, Y.}, \bibinfo{author}{Geng, B.}, \bibinfo{author}{Tang,
  B.}, \bibinfo{year}{2023}.
\newblock \bibinfo{title}{Simplified coded dispersion entropy: A nonlinear
  metric for signal analysis}.
\newblock \bibinfo{journal}{Nonlinear Dynamics} \bibinfo{volume}{111},
  \bibinfo{pages}{9327--9344}.
\bibitem[{Li et~al.(2022c)Li, Tang, Geng and Jiao}]{bear3}
\bibinfo{author}{Li, Y.}, \bibinfo{author}{Tang, B.}, \bibinfo{author}{Geng,
  B.}, \bibinfo{author}{Jiao, S.}, \bibinfo{year}{2022}c.
\newblock \bibinfo{title}{Fractional order fuzzy dispersion entropy and its
  application in bearing fault diagnosis}.
\newblock \bibinfo{journal}{fractal and fractional} \bibinfo{volume}{6},
  \bibinfo{pages}{544}.
\bibitem[{Lu et~al.(2022)Lu, Fu, Yue, Zhu, Wang and Hu}]{nd5}
\bibinfo{author}{Lu, J.}, \bibinfo{author}{Fu, Y.}, \bibinfo{author}{Yue, J.},
  \bibinfo{author}{Zhu, L.}, \bibinfo{author}{Wang, D.}, \bibinfo{author}{Hu,
  Z.}, \bibinfo{year}{2022}.
\newblock \bibinfo{title}{Natural gas pipeline leak diagnosis based on improved
  variational modal decomposition and locally linear embedding feature
  extraction method}.
\newblock \bibinfo{journal}{Process Safety and Environmental Protection}
  \bibinfo{volume}{164}, \bibinfo{pages}{857--867}.
\bibitem[{Mallat(1989)}]{wavelet}
\bibinfo{author}{Mallat, S.G.}, \bibinfo{year}{1989}.
\newblock \bibinfo{title}{A theory for multiresolution signal decomposition:
  the wavelet representation}.
\newblock \bibinfo{journal}{IEEE transactions on pattern analysis and machine
  intelligence} \bibinfo{volume}{11}, \bibinfo{pages}{674--693}.
\bibitem[{Pincus(1991)}]{AE}
\bibinfo{author}{Pincus, S.M.}, \bibinfo{year}{1991}.
\newblock \bibinfo{title}{Approximate entropy as a measure of system
  complexity.}
\newblock \bibinfo{journal}{Proceedings of the national academy of sciences}
  \bibinfo{volume}{88}, \bibinfo{pages}{2297--2301}.
\bibitem[{Richman et~al.(2004)Richman, Lake and Moorman}]{SE}
\bibinfo{author}{Richman, J.S.}, \bibinfo{author}{Lake, D.E.},
  \bibinfo{author}{Moorman, J.R.}, \bibinfo{year}{2004}.
\newblock \bibinfo{title}{Sample entropy}, in: \bibinfo{booktitle}{Methods in
  enzymology}. \bibinfo{publisher}{Elsevier}. volume \bibinfo{volume}{384}, pp.
  \bibinfo{pages}{172--184}.
\bibitem[{Rostaghi and Azami(2016)}]{DE}
\bibinfo{author}{Rostaghi, M.}, \bibinfo{author}{Azami, H.},
  \bibinfo{year}{2016}.
\newblock \bibinfo{title}{Dispersion entropy: A measure for time-series
  analysis}.
\newblock \bibinfo{journal}{IEEE Signal Processing Letters}
  \bibinfo{volume}{23}, \bibinfo{pages}{610--614}.
\bibitem[{Rostaghi et~al.(2021)Rostaghi, Khatibi, Ashory and Azami}]{DE3}
\bibinfo{author}{Rostaghi, M.}, \bibinfo{author}{Khatibi, M.M.},
  \bibinfo{author}{Ashory, M.R.}, \bibinfo{author}{Azami, H.},
  \bibinfo{year}{2021}.
\newblock \bibinfo{title}{Fuzzy dispersion entropy: A nonlinear measure for
  signal analysis}.
\newblock \bibinfo{journal}{IEEE Transactions on Fuzzy Systems}
  \bibinfo{volume}{30}, \bibinfo{pages}{3785--3796}.
\bibitem[{Shang and Shang(2024)}]{nd1}
\bibinfo{author}{Shang, B.}, \bibinfo{author}{Shang, P.}, \bibinfo{year}{2024}.
\newblock \bibinfo{title}{A novel and effective method for quantifying
  complexity of nonlinear time series}.
\newblock \bibinfo{journal}{Communications in Nonlinear Science and Numerical
  Simulation} \bibinfo{volume}{130}, \bibinfo{pages}{107773}.
\bibitem[{Shannon(1948)}]{shannon}
\bibinfo{author}{Shannon, C.E.}, \bibinfo{year}{1948}.
\newblock \bibinfo{title}{A mathematical theory of communication}.
\newblock \bibinfo{journal}{The Bell system technical journal}
  \bibinfo{volume}{27}, \bibinfo{pages}{379--423}.
\bibitem[{Small(2013)}]{OPTN}
\bibinfo{author}{Small, M.}, \bibinfo{year}{2013}.
\newblock \bibinfo{title}{Complex networks from time series: Capturing
  dynamics}, in: \bibinfo{booktitle}{2013 IEEE International Symposium on
  Circuits and Systems (ISCAS)}, \bibinfo{organization}{IEEE}. pp.
  \bibinfo{pages}{2509--2512}.
\bibitem[{Smith and Randall(2015)}]{CWRU}
\bibinfo{author}{Smith, W.A.}, \bibinfo{author}{Randall, R.B.},
  \bibinfo{year}{2015}.
\newblock \bibinfo{title}{Rolling element bearing diagnostics using the case
  western reserve university data: A benchmark study}.
\newblock \bibinfo{journal}{Mechanical Systems and Signal Processing}
  \bibinfo{volume}{64-65}, \bibinfo{pages}{100--131}.
\newblock \DOIprefix\doi{https://doi.org/10.1016/j.ymssp.2015.04.021}.
\bibitem[{Wang et~al.(2023)Wang, Shang and Shang}]{nd2}
\bibinfo{author}{Wang, Z.}, \bibinfo{author}{Shang, P.},
  \bibinfo{author}{Shang, B.}, \bibinfo{year}{2023}.
\newblock \bibinfo{title}{Time irreversibility analysis and abnormality
  detection based on riemannian geometry for complex time series}.
\newblock \bibinfo{journal}{Communications in Nonlinear Science and Numerical
  Simulation} \bibinfo{volume}{117}, \bibinfo{pages}{106985}.
\bibitem[{Wang et~al.(2020)Wang, Yao and Cai}]{bear2}
\bibinfo{author}{Wang, Z.}, \bibinfo{author}{Yao, L.}, \bibinfo{author}{Cai,
  Y.}, \bibinfo{year}{2020}.
\newblock \bibinfo{title}{Rolling bearing fault diagnosis using generalized
  refined composite multiscale sample entropy and optimized support vector
  machine}.
\newblock \bibinfo{journal}{Measurement} \bibinfo{volume}{156},
  \bibinfo{pages}{107574}.
\bibitem[{Yan and Jia(2019)}]{DE2}
\bibinfo{author}{Yan, X.}, \bibinfo{author}{Jia, M.}, \bibinfo{year}{2019}.
\newblock \bibinfo{title}{Intelligent fault diagnosis of rotating machinery
  using improved multiscale dispersion entropy and mrmr feature selection}.
\newblock \bibinfo{journal}{Knowledge-Based Systems} \bibinfo{volume}{163},
  \bibinfo{pages}{450--471}.
\bibitem[{Zhang and Shang(2021)}]{TPE}
\bibinfo{author}{Zhang, B.}, \bibinfo{author}{Shang, P.}, \bibinfo{year}{2021}.
\newblock \bibinfo{title}{Transition permutation entropy and transition
  dissimilarity measure: Efficient tools for fault detection of railway vehicle
  systems}.
\newblock \bibinfo{journal}{IEEE transactions on industrial informatics}
  \bibinfo{volume}{18}, \bibinfo{pages}{1654--1662}.
\bibitem[{Zhao et~al.(2023)Zhao, Chen, Gui, Liu and Yang}]{Zhao}
\bibinfo{author}{Zhao, Z.}, \bibinfo{author}{Chen, F.}, \bibinfo{author}{Gui,
  Z.}, \bibinfo{author}{Liu, D.}, \bibinfo{author}{Yang, J.},
  \bibinfo{year}{2023}.
\newblock \bibinfo{title}{Refined composite hierarchical multiscale lempel-ziv
  complexity: A quantitative diagnostic method of multi-feature fusion for
  rotating energy devices}.
\newblock \bibinfo{journal}{Renewable Energy} \bibinfo{volume}{218},
  \bibinfo{pages}{119310}.
\bibitem[{Zozor et~al.(2005)Zozor, Ravier and Buttelli}]{LZC2}
\bibinfo{author}{Zozor, S.}, \bibinfo{author}{Ravier, P.},
  \bibinfo{author}{Buttelli, O.}, \bibinfo{year}{2005}.
\newblock \bibinfo{title}{On lempel--ziv complexity for multidimensional data
  analysis}.
\newblock \bibinfo{journal}{Physica A: Statistical Mechanics and its
  Applications} \bibinfo{volume}{345}, \bibinfo{pages}{285--302}.

\end{thebibliography}

\bio{}
\endbio

\endbio

\end{document}